\RequirePackage{lineno}

\documentclass[aps,prd,twocolumn,showpacs,superscriptaddress,groupedaddress,letter]{revtex4}  
\usepackage{graphicx}  
\usepackage{dcolumn}   
\usepackage{bm}        
\usepackage{amssymb}   

\usepackage{mathptmx}
\usepackage{lineno}
\usepackage{xspace}
\usepackage{multirow}
\usepackage{epsfig}
\usepackage{psfrag}
\usepackage{lscape}
\usepackage{comment}

\newcommand{\dzero}{D0\xspace}

\newcommand{\runii}{Run~II\xspace}

\newcommand{\ppbar}{\ensuremath{p\bar{p}}\xspace}
\newcommand{\ttbar}{\ensuremath{t\bar{t}}\xspace}
\newcommand{\etmiss}{\ensuremath{E \kern-0.6em\slash_{\rm T}}\xspace}
\newcommand{\etmissx}{\ensuremath{E \kern-0.6em\slash_{\rm x}}\xspace}
\newcommand{\etmissy}{\ensuremath{E \kern-0.6em\slash_{\rm y}}\xspace}
\newcommand{\alpgen}{{\sc alpgen}\xspace}
\newcommand{\pythia}{{\sc pythia}\xspace}
\newcommand{\herwig}{{\sc Herwig}\xspace}
\newcommand{\vecbos}{{\sc vecbos}\xspace}

\newcommand{\dm}{\ensuremath{\Delta m}\xspace}
\newcommand{\ddm}{\ensuremath{\delta_{\Delta m}}\xspace}
\newcommand{\msum}{\ensuremath{m_{\rm top}}\xspace}
\newcommand{\mt}{\ensuremath{m_t}\xspace}
\newcommand{\mtb}{\ensuremath{m_{\bar t}}\xspace}
\newcommand{\ejets}{\ensuremath{e\!+\!{\rm jets}}\xspace}
\newcommand{\mujets}{\ensuremath{\mu\!+\!{\rm jets}}\xspace}
\newcommand{\ljets}{\ensuremath{\ell\!+\!{\rm jets}}\xspace}
\newcommand{\wjets}{\ensuremath{W\!+\!{\rm jets}}\xspace}
\newcommand{\zjets}{\ensuremath{Z\!+\!{\rm jets}}\xspace}
\newcommand{\jets}{\ensuremath{\rm jets}\xspace}
\newcommand{\fsig}{\ensuremath{f}\xspace}
\newcommand{\pevt}{\ensuremath{P_{\rm evt}}\xspace}
\newcommand{\psig}{\ensuremath{P_{\rm sig}}\xspace}
\newcommand{\pbkg}{\ensuremath{P_{\rm bkg}}\xspace}
\newcommand{\acc}{\ensuremath{A}\xspace}
\newcommand{\etj}{\ensuremath{E_T^j}\xspace}
\newcommand{\rbb}{\ensuremath{\mathcal{R}_{b,\bar b}}\xspace}
\newcommand{\pullw}[1]{\ensuremath{\omega_{\pi_{#1}}}\xspace}

\newcommand{\eps}{\varepsilon}

\newcommand{\etal}{\textit{et~al.}}
\newcommand{\mum}{\ensuremath{\mu\textnormal{m}}\xspace}

\newcommand{\GeV}{\ensuremath{\textnormal{GeV}}\xspace}

\newcommand{\dif}{\ensuremath{{\rm d}}}

\newcommand{\met}{\ensuremath{p\!\!\!\!/_T}\xspace}

\newcommand{\fb}{\ensuremath{{\rm fb}^{-1}}\xspace}
\newcommand{\mtop}{\ensuremath{m_{\rm top}}\xspace}

\newcommand{\pt}{\ensuremath{p_T}\xspace}

\newcommand{\jes}{\ensuremath{k_{\rm JES}}\xspace}
\newcommand{\etat}{\ensuremath{\tilde\eta}\xspace}

\begin{document}

\hspace{5.2in} \mbox{Fermilab-Pub-11/257-E}

\title{Direct measurement of the mass difference between top and antitop quarks}
\affiliation{Universidad de Buenos Aires, Buenos Aires, Argentina}
\affiliation{LAFEX, Centro Brasileiro de Pesquisas F{\'\i}sicas, Rio de Janeiro, Brazil}
\affiliation{Universidade do Estado do Rio de Janeiro, Rio de Janeiro, Brazil}
\affiliation{Universidade Federal do ABC, Santo Andr\'e, Brazil}
\affiliation{Instituto de F\'{\i}sica Te\'orica, Universidade Estadual Paulista, S\~ao Paulo, Brazil}
\affiliation{Simon Fraser University, Vancouver, British Columbia, and York University, Toronto, Ontario, Canada}
\affiliation{University of Science and Technology of China, Hefei, People's Republic of China}
\affiliation{Universidad de los Andes, Bogot\'{a}, Colombia}
\affiliation{Charles University, Faculty of Mathematics and Physics, Center for Particle Physics, Prague, Czech Republic}
\affiliation{Czech Technical University in Prague, Prague, Czech Republic}
\affiliation{Center for Particle Physics, Institute of Physics, Academy of Sciences of the Czech Republic, Prague, Czech Republic}
\affiliation{Universidad San Francisco de Quito, Quito, Ecuador}
\affiliation{LPC, Universit\'e Blaise Pascal, CNRS/IN2P3, Clermont, France}
\affiliation{LPSC, Universit\'e Joseph Fourier Grenoble 1, CNRS/IN2P3, Institut National Polytechnique de Grenoble, Grenoble, France}
\affiliation{CPPM, Aix-Marseille Universit\'e, CNRS/IN2P3, Marseille, France}
\affiliation{LAL, Universit\'e Paris-Sud, CNRS/IN2P3, Orsay, France}
\affiliation{LPNHE, Universit\'es Paris VI and VII, CNRS/IN2P3, Paris, France}
\affiliation{CEA, Irfu, SPP, Saclay, France}
\affiliation{IPHC, Universit\'e de Strasbourg, CNRS/IN2P3, Strasbourg, France}
\affiliation{IPNL, Universit\'e Lyon 1, CNRS/IN2P3, Villeurbanne, France and Universit\'e de Lyon, Lyon, France}
\affiliation{III. Physikalisches Institut A, RWTH Aachen University, Aachen, Germany}
\affiliation{Physikalisches Institut, Universit{\"a}t Freiburg, Freiburg, Germany}
\affiliation{II. Physikalisches Institut, Georg-August-Universit{\"a}t G\"ottingen, G\"ottingen, Germany}
\affiliation{Institut f{\"u}r Physik, Universit{\"a}t Mainz, Mainz, Germany}
\affiliation{Ludwig-Maximilians-Universit{\"a}t M{\"u}nchen, M{\"u}nchen, Germany}
\affiliation{Fachbereich Physik, Bergische Universit{\"a}t Wuppertal, Wuppertal, Germany}
\affiliation{Panjab University, Chandigarh, India}
\affiliation{Delhi University, Delhi, India}
\affiliation{Tata Institute of Fundamental Research, Mumbai, India}
\affiliation{University College Dublin, Dublin, Ireland}
\affiliation{Korea Detector Laboratory, Korea University, Seoul, Korea}
\affiliation{CINVESTAV, Mexico City, Mexico}
\affiliation{Nikhef, Science Park, Amsterdam, the Netherlands}
\affiliation{Radboud University Nijmegen, Nijmegen, the Netherlands and Nikhef, Science Park, Amsterdam, the Netherlands}
\affiliation{Joint Institute for Nuclear Research, Dubna, Russia}
\affiliation{Institute for Theoretical and Experimental Physics, Moscow, Russia}
\affiliation{Moscow State University, Moscow, Russia}
\affiliation{Institute for High Energy Physics, Protvino, Russia}
\affiliation{Petersburg Nuclear Physics Institute, St. Petersburg, Russia}
\affiliation{Instituci\'{o} Catalana de Recerca i Estudis Avan\c{c}ats (ICREA) and Institut de F\'{i}sica d'Altes Energies (IFAE), Barcelona, Spain}
\affiliation{Stockholm University, Stockholm and Uppsala University, Uppsala, Sweden}
\affiliation{Lancaster University, Lancaster LA1 4YB, United Kingdom}
\affiliation{Imperial College London, London SW7 2AZ, United Kingdom}
\affiliation{The University of Manchester, Manchester M13 9PL, United Kingdom}
\affiliation{University of Arizona, Tucson, Arizona 85721, USA}
\affiliation{University of California Riverside, Riverside, California 92521, USA}
\affiliation{Florida State University, Tallahassee, Florida 32306, USA}
\affiliation{Fermi National Accelerator Laboratory, Batavia, Illinois 60510, USA}
\affiliation{University of Illinois at Chicago, Chicago, Illinois 60607, USA}
\affiliation{Northern Illinois University, DeKalb, Illinois 60115, USA}
\affiliation{Northwestern University, Evanston, Illinois 60208, USA}
\affiliation{Indiana University, Bloomington, Indiana 47405, USA}
\affiliation{Purdue University Calumet, Hammond, Indiana 46323, USA}
\affiliation{University of Notre Dame, Notre Dame, Indiana 46556, USA}
\affiliation{Iowa State University, Ames, Iowa 50011, USA}
\affiliation{University of Kansas, Lawrence, Kansas 66045, USA}
\affiliation{Kansas State University, Manhattan, Kansas 66506, USA}
\affiliation{Louisiana Tech University, Ruston, Louisiana 71272, USA}
\affiliation{Boston University, Boston, Massachusetts 02215, USA}
\affiliation{Northeastern University, Boston, Massachusetts 02115, USA}
\affiliation{University of Michigan, Ann Arbor, Michigan 48109, USA}
\affiliation{Michigan State University, East Lansing, Michigan 48824, USA}
\affiliation{University of Mississippi, University, Mississippi 38677, USA}
\affiliation{University of Nebraska, Lincoln, Nebraska 68588, USA}
\affiliation{Rutgers University, Piscataway, New Jersey 08855, USA}
\affiliation{Princeton University, Princeton, New Jersey 08544, USA}
\affiliation{State University of New York, Buffalo, New York 14260, USA}
\affiliation{Columbia University, New York, New York 10027, USA}
\affiliation{University of Rochester, Rochester, New York 14627, USA}
\affiliation{State University of New York, Stony Brook, New York 11794, USA}
\affiliation{Brookhaven National Laboratory, Upton, New York 11973, USA}
\affiliation{Langston University, Langston, Oklahoma 73050, USA}
\affiliation{University of Oklahoma, Norman, Oklahoma 73019, USA}
\affiliation{Oklahoma State University, Stillwater, Oklahoma 74078, USA}
\affiliation{Brown University, Providence, Rhode Island 02912, USA}
\affiliation{University of Texas, Arlington, Texas 76019, USA}
\affiliation{Southern Methodist University, Dallas, Texas 75275, USA}
\affiliation{Rice University, Houston, Texas 77005, USA}
\affiliation{University of Virginia, Charlottesville, Virginia 22901, USA}
\affiliation{University of Washington, Seattle, Washington 98195, USA}
\author{V.M.~Abazov} \affiliation{Joint Institute for Nuclear Research, Dubna, Russia}
\author{B.~Abbott} \affiliation{University of Oklahoma, Norman, Oklahoma 73019, USA}
\author{B.S.~Acharya} \affiliation{Tata Institute of Fundamental Research, Mumbai, India}
\author{M.~Adams} \affiliation{University of Illinois at Chicago, Chicago, Illinois 60607, USA}
\author{T.~Adams} \affiliation{Florida State University, Tallahassee, Florida 32306, USA}
\author{G.D.~Alexeev} \affiliation{Joint Institute for Nuclear Research, Dubna, Russia}
\author{G.~Alkhazov} \affiliation{Petersburg Nuclear Physics Institute, St. Petersburg, Russia}
\author{A.~Alton$^{a}$} \affiliation{University of Michigan, Ann Arbor, Michigan 48109, USA}
\author{G.~Alverson} \affiliation{Northeastern University, Boston, Massachusetts 02115, USA}
\author{G.A.~Alves} \affiliation{LAFEX, Centro Brasileiro de Pesquisas F{\'\i}sicas, Rio de Janeiro, Brazil}
% ...
%
%
\collaboration{The D0 Collaboration\footnote{with visitors from
$^{a}$Augustana College, Sioux Falls, SD, USA,
$^{b}$The University of Liverpool, Liverpool, UK,
$^{c}$SLAC, Menlo Park, CA, USA,
$^{d}$University College London, London, UK,
$^{e}$Centro de Investigacion en Computacion - IPN, Mexico City, Mexico,
$^{f}$ECFM, Universidad Autonoma de Sinaloa, Culiac\'an, Mexico,
and 
$^{g}$Universit{\"a}t Bern, Bern, Switzerland.
$^{\ddag}$Deceased.
}} \noaffiliation
\vskip 0.25cm
\date{June 10, 2011}

\begin{abstract}
We present a direct measurement of the mass difference between top and antitop quarks (\dm) in lepton$+$jets \ttbar final states using the ``matrix element'' method. 
The purity of the lepton$+$jets sample is enhanced for \ttbar events by identifying at least one of the jet as originating from a $b$ quark.
The analyzed data correspond to 3.6\,fb$^{-1}$ of $\ppbar$ collisions at $\sqrt s = 1.96~{\rm TeV}$ acquired
by \dzero in Run~II of the Fermilab Tevatron Collider.
The combination of the \ejets and \mujets channels yields $\dm=0.8\pm 1.8~(\mbox{stat)}\pm 0.5~(\mbox{syst)~\GeV}$, which is in agreement with the standard model expectation of no mass difference.
%
\end{abstract}

\maketitle


\section{Introduction}
The standard model (SM) is a local gauge-invariant quantum field theory (QFT), with invariance under charge, parity, and time reversal ($CPT$) providing one of its most fundamental principles~\cite{bib:cpt_qft1,bib:cpt_qft2,bib:cpt_qft3,bib:cpt_qft4}, which also constrains the SM~\cite{bib:cpt_sm}.
In fact, any Lorentz-invariant local QFT must conserve $CPT$~\cite{bib:cpt_local}. A difference in the mass of a particle and its antiparticle would constitute a violation of $CPT$ invariance. This issue has been tested extensively for many elementary particles of the SM~\cite{bib:pdg}. Quarks, however, carry color charge, and therefore are not observed directly, but must first hadronize via quantum chromodynamic (QCD) processes into jets of colorless particles. These hadronization products reflect properties of the initially produced quarks, such as their masses, electric charges, and spin states. Except for the top quark, the time scale for hadronization of quarks is orders of magnitude less than for electroweak decay, thereby favoring the formation of QCD-bound hadronic states before decay. This introduces a significant dependence of the mass of a quark on the model of QCD binding and evolution. In contrast to other quarks, no bound states are formed before decay of produced top quarks, thereby providing a unique opportunity to measure directly the mass difference between a quark and its antiquark~\cite{bib:dmtheory}.

In proton-antiproton collisions at the Fermilab Tevatron Collider, top quarks are produced in $\ttbar$ pairs via the strong interaction, or singly via the electroweak interaction.
In the SM, the top quark decays almost exclusively into a $W$~boson and a $b$~quark. The topology of a $\ttbar$~event is therefore determined by the subsequent decays of the $W$~bosons. The world's most precise top quark mass measurements are performed in the lepton$+$jets (\ljets) channels, which are characterized by the presence of one isolated energetic electron or muon from one $W\to\ell\nu$ decay, an imbalance in transverse momentum relative to the beam axis from the escaping neutrino, and four or more jets from the evolution of the two $b$~quarks and the two quarks from the second $W\to q\bar q'$ decay.

The top quark was discovered~\cite{bib:discovery1,bib:discovery2} in proton-antiproton collision data at a center of mass energy of $\sqrt s=1.8~$TeV in  Run~I of the Tevatron. After an upgrade to a higher center of mass energy of $\sqrt s=1.96~$TeV and higher luminosities, Run~II of the Tevatron commenced in~2001. Since then, a large sample of \ttbar\ events has been collected, yielding precision measurements of various SM parameters such as the mass of the top quark, which has been determined to an accuracy of about $0.6\%$ or $\mtop\equiv\frac12(\mt+\mtb) = 173.3\pm1.1~\GeV$~\cite{bib:combi}, where \mt~(\mtb) is the mass of the top~(antitop) quark.

The \dzero\ Collaboration published the first measurement of the top-antitop quark mass difference, $\dm\equiv \mt-\mtb$, using 1\,fb$^{-1}$ of Run~II integrated luminosity~\cite{bib:p17dm}. Our new measurement, presented here, employs the same matrix element (ME) technique~\cite{Aba06,Dem08}, suggested initially by Kondo~\etal~\cite{bib:me1,bib:me2,bib:me3}, and developed to its current form by~\dzero~\cite{Aba04}. Our previous study measured a mass difference
  \[\dm=3.8 \pm 3.4\,(\rm stat.) \pm 1.2\,(\rm syst.)~\GeV.\]
Recently, CDF has also measured \dm~\cite{bib:dmcdf} based on 5.6\,\fb\ of \runii data, using a template technique, and found
  \[\dm=-3.3 \pm 1.4\,(\rm stat.) \pm 1.0\,(\rm syst.)~\GeV.\]

In this paper, we extend our first measurement of \dm using an additional 2.6\,\fb\ of Run~II integrated luminosity, and combining our two results. 
We also re-examine the uncertainties from the modeling of signal processes and of the response of the detector. 
Most important is a possible presence of asymmetries in the calorimeter response to $b$ and $\bar b$-quark jets, which we re-evaluate using a purely data-driven method. We also consider for the first time a bias from asymmetries in response to $c$ and $\bar c$-quark jets. 

This paper is arranged as follows: after a brief descripton of the \dzero detector in Sec.~\ref{sec:detector}, we review the event selection and reconstruction in Sec.~\ref{sec:selection}. In Sec.~\ref{sec:samples}, we define the samples of Monte Carlo (MC) events used in the analysis.  The extraction of the top-antitop quark mass difference using the ME technique is then briefly reviewed in Sec.~\ref{sec:method}. The calibration of this technique, based on MC events, and the measurement of the mass difference in 2.6\,\fb\ of Run II integrated luminosity are presented in Sec.~\ref{sec:measurement}. The evaluation of systematic uncertainties and cross checks are discussed in Sec.~\ref{sec:syst} and~\ref{sec:cross}, respectively. Finally, the combination of the measurements for the 2.6\,\fb\ and 1\,\fb\ data samples is presented in Sec.~\ref{sec:combi}. 

\section{The \dzero\ detector\label{sec:detector}}
The \dzero detector has a central-tracking system, calorimetry, and a muon system. 
The central-tracking system consists of a 
silicon microstrip tracker (SMT) and a central fiber tracker (CFT), 
both located within a 1.9\,T superconducting solenoidal 
magnet~\cite{d0det,bib:layer0,bib:layer0disclaimer}, with designs optimized for tracking and 
vertexing at pseudorapidities $|\eta|<3$~\cite{eta}. The SMT can reconstruct the
$p\bar p$ interaction vertex (PV) with a precision of about 40\,$\mum$ in the
plane transverse to the beam direction and determine the impact
parameter of any track relative to the PV~\cite{ip} with a precision between 
20 and 50\,$\mum$, depending on the number of hits in the SMT.  
These are the key elements to lifetime-based $b$-quark jet tagging.
The liquid-argon and uranium sampling calorimeter has a 
central section (CC) covering pseudorapidities $|\eta|\lesssim1.1$ and two end calorimeters (EC) that extend coverage 
to $|\eta|\approx 4.2$, with all three housed in separate 
cryostats~\cite{d0det,run1det}. Central and forward preshower detectors are positioned just before the CC and EC. An outer muon system, at $|\eta|<2$, 
consists of a layer of tracking detectors and scintillation trigger 
counters in front of 1.8\,T toroids, followed by two similar layers 
after the toroids~\cite{run2muon}. 
The luminosity is calculated from the rate of 
$p \bar p$ inelastic collisions measured with plastic
scintillator arrays, which are located in front of the EC
cryostats. The trigger and data 
acquisition systems are designed to accommodate the high instantaneous luminosities 
of Run~II~\cite{bib:l1cal2b}.

\section{Event selection\label{sec:selection}}
In this new measurement of \dm, we analyze data 
corresponding to an integrated luminosity  of about $2.6\,\fb$ for both the \ejets and \mujets channels.

Candidate $\ttbar$ events are required to pass an isolated energetic lepton trigger or a lepton$+$jet(s) trigger. 
These events are enriched in $\ttbar$ content by requiring exactly four jets reconstructed using the Run~II cone algorithm~\cite{RunIIcone} with cone radius $\Delta R\equiv\sqrt{(\Delta\eta)^2+(\Delta\phi)^2}=0.5$, transverse momenta $\pt>20$~\GeV, and pseudorapidities $|\eta|<2.5$. The jet of highest transverse momentum in a given event must have $\pt > 40$~\GeV. Furthermore, we require exactly one isolated electron with $\pt>20$~\GeV and $|\eta|<1.1$, or exactly one isolated muon with $\pt>20$~\GeV and $|\eta|<2.0$. 
The leptons must originate within 1\,cm of the PV in the coordinate along the beamline.
Events containing an additional isolated lepton (either $e$ or~$\mu$) with $\pt>15$~\GeV are rejected. Lepton isolation criteria are based on calorimetric and tracking  information along with object identification criteria, as described in Ref.~\cite{PRDtopo}. The positively (negatively) charged leptons are used to tag the top (antitop) quark in a given event. To reduce instrumental effects that can cause charge-dependent asymmetries in the lepton momentum scale, the polarity of the solenoidal magnetic field is routinely reversed, splitting the total data into two samples of approximately equal size. The PV must have at least three associated tracks and lie 
within the fiducial region of the SMT. At least one neutrino is expected in the \ljets final state; hence, an imbalance in transverse momentum (defined as the opposite of the vector sum of the transverse energies in each calorimeter cell, corrected for the energy carried by identified muons and energy added or subtracted due to the jet energy scale calibration described below) of $\met>20~\GeV~(25~\GeV)$ must be present in the \ejets\ (\mujets) channel. These kinematic selections are summarized in Table~\ref{tab:selection}. 

To reduce the contribution of multijet production (MJ) in the \ejets channel, $\Delta\phi(e,\met) > 2.2-\met\times0.045~\GeV^{-1}$ is required for the azimuthal difference $\Delta\phi(e,\met)=|\phi_{e}-\phi_{\met}|$ between the electron and the direction of~$\met$. Likewise, $\Delta\phi(\mu,\met) > 2.1 - \met\times0.035~\GeV^{-1}$ is required in the \mujets channel.
Jets from \mbox{$b$ quarks} are identified by a neural-network-based $b$-tagging algorithm~\cite{btagging}, which combines variables that characterize properties of secondary vertices and tracks within the jet that have large impact parameters relative to the PV. Typically, its efficiency for $b$-quark jets is about 65\%, while the probability for misidentifying $u$, $d$, $s$-quark and gluon jets as $b$ jets is about 3\%.
To increase \ttbar purity, and to reduce the number of combinatoric possibilities for assigning jets to \ttbar decay products, we require at least one $b$-tagged jet to be present in the events used to measure $\dm$. 

After all acceptance requirements, a data sample of 312 (303) events is selected in the \ejets (\mujets) channel. As discussed above, each of those samples is split according to lepton charge. In the \ejets channel, 174~(138) events have a positive (negative) lepton in the final state. Likewise, the \mujets sample is split to subsets of 145 and 158 events.

\begin{table}
\caption{\label{tab:selection}
A summary of kinematic event selections applied.
}
\begin{ruledtabular}
\begin{tabular}{lll}
\multirow{2}{*}{Exactly 1 charged lepton}
                          & $\pt > 20~\GeV$ & $|\eta| < 1.1$ ($e$)\\
                          & $\pt > 20~\GeV$ & $|\eta| < 2.0$ ($\mu$)\\
Exactly 4 jets            & $\pt > 20~\GeV$ & $|\eta| < 2.5$\\
Jet of highest \pt        & $\pt > 40~\GeV$ & $|\eta| < 2.5$\\
\multirow{2}{*}{Imbalance in transverse momentum}
                          & $\met > 20\,~\GeV$ & (\ejets) \\
                          & $\met > 25\,~\GeV$ & (\mujets) \\
\end{tabular}
\end{ruledtabular}
\end{table}

\section{Monte carlo simulation\label{sec:samples}}
Large samples of simulated MC events are used to determine the resolution of the detector and to calibrate the \dm measurement as well as the statistical sensitivity of the method. After simulation of the hard scattering part of the interaction and parton shower corrections, MC events are passed through a detailed detector simulation based on {\sc geant}~\cite{bib:geant}, overlaid with data collected from a random subsample of beam crossings to model the effects of noise and multiple interactions, and reconstructed using the same algorithms that are used for data.
Although the fraction of signal events, $\fsig$, is fitted in the analysis, we also cross check that the entire data sample is described adequately by the simulations.

\subsection{Monte Carlo samples for signal\label{ssec:samplessig}}
Simulated \ttbar events with different \mt and \mtb are required to calibrate the \dm measurement. We use the \pythia generator~\cite{bib:pythia}, version 6.413, to model the \ttbar signal. This generator models the Breit-Wigner shape of the invariant mass distribution of $t$ and $\bar t$ quarks, whose correct description is important for the \dm measurement. 

In the standard \pythia, it is not possible to generate \ttbar events with different masses $\mt$ and $\mtb$. Therefore, we modify the \pythia program to provide signal events with $\mt\neq\mtb$. In applying these modifications, we adjust the description of all quantities that depend on the two masses, for example, the respective decay widths $\Gamma_t$ and $\Gamma_{\bar t}$. Technical details of this implementation can be found in Appendix~\ref{sec:app}.

We generate \ttbar events using the CTEQ6L1 parton distribution function set (PDF)~\cite{bib:cteq} at the momentum transfer scale $Q^2 = (\pt^{\rm scat})^2 + \frac{1}{2}\left\{ P_{1}^2 + P_{2}^2 + \mt^{2} + \mtb^{2} \right\}$, where $\pt^{\rm scat}$ is the transverse momentum for the hard scattering process, and $P_{i}$ is the four-momentum of the incoming parton $i$. For $\mt=\mtb$, the expression used for $Q^2$ is identical to that in the standard \pythia. All other steps in the event simulation process aside from the generation of the hard-scattering process, e.g., the modeling of the detector response, are unchanged from the standard \pythia.  

We check our modified \pythia version against the original by comparing large samples of simulated $\ttbar$ events for $(\mt,\mtb)=(170~\GeV,\,170~\GeV)$, at both the parton and reconstruction levels, and find full consistency. 


The $\ttbar$ samples are generated at fourteen combinations of top and antitop quark masses $(\mt,\mtb)$, which form a grid spaced at 5~\GeV intervals between (165~\GeV,\,165~\GeV) and (180~\GeV,\,180~\GeV), excluding the two extreme points at (165~\GeV,\,180~\GeV) and (180~\GeV,\,165~\GeV). 
The four points with $\mt=\mtb$ are generated with the standard \pythia, whereas all others use our modified version of the generator.

\subsection{Monte Carlo and other simulations of background}
The dominant background to \ttbar decays into $\ljets$ final states is from the electroweak production of a~$W$~boson in association with jets from gluon radiation. We simulate the hard scattering part of this process
using the \alpgen MC program~\cite{bib:alpgen}, which is capable of simulating up to five additional particles in the final state at leading order (LO) in $\alpha_s$.
\alpgen is coupled to \pythia, which is used to model the hadronization of the partons and the evolution of the shower. The MLM matching scheme is applied to avoid double-counting of partonic event configurations~\cite{bib:matching}. The \wjets contribution is divided into two categories according to parton flavor: 
$(i)$~$W\!+\!b\bar b\!+\!\jets$ and $W\!\!+\!c\bar c\!+\!\jets$ and (ii) all other contributions, where ``jets'' generically denotes jets from $u,d,s$-quarks and gluons. The second category also includes the $W\!\!+\!c\!+\!\jets$ final states.
While the individual processes are generated with \alpgen, the relative contributions of the two categories are determined using next-to-LO (NLO) calculations, with next-to-leading logarithmic (NLL) corrections based on the {\sc mcfm} MC generator~\cite{bib:mcfm}. These NLO corrections increase the LO cross section of category $(i)$ by a factor of $k=1.47\pm0.22$, while $k=1$ is used for category $(ii)$. The resulting combined $\wjets$ background contribution is then determined from a fit to data and predictions for other signal and background contributions, as described in Sec.~\ref{sec:method}. Thus, the NLO $k$-factors only change the relative balance between $(i)$ and $(ii)$.

Additional background contributions arise from $WW$, $WZ$, $ZZ$, single top quark electroweak production, $Z\rightarrow\tau\tau$, and $Z\rightarrow ee$ ($Z\rightarrow \mu\mu$) production in the \ejets (\mujets) channel. The predictions for these backgrounds are taken from MC simulations, and, with the exception of single top quark electroweak production, their production cross sections are normalized to NLO$+$NLL calculations with {\sc mcfm}. Diboson processes are simulated with \pythia. The hard-scattering part of single top quark production is simulated with {\sc CompHEP}~\cite{bib:comphep}, while \alpgen is used for $Z\!+\!\jets$ boson production. For both backgrounds, \pythia is employed to model hadronization and shower evolution. The CTEQ6L1 PDFs and the D0~Tune~A underlying event model~\cite{bib:d0tunea} are used in the generation of all MC samples.

Events from MJ production can pass our selection criteria, which typically happens when a jets mimics an electron, or a muon that arises from a semileptonic decay of a $b$ or $c$ quark appears to be isolated. The kinematic distributions of the MJ background are modeled using events in data that fail only the electron identification (muon isolation) criteria, but pass loosened versions of these criteria defined in~\cite{bib:matrixmethod}. The absolute contribution of this background to each of the channels is estimated using the method described in Ref.~\cite{bib:matrixmethod}. This method uses the absolute numbers of events with prompt leptons $N^{\ttbar+W}_{\rm loose}$ and events from MJ production $N^{\rm MJ}_{\rm loose}$ in the sample with loosened lepton identification criteria, and relates them to the absolute contributions to the sample with standard lepton identification criteria via $N=\eps^{\ttbar+W}N^{\ttbar+W}_{\rm loose}+\eps^{\rm MJ}N^{\rm MJ}_{\rm loose}$. Here, $\eps^{\ttbar+W}$ and $\eps^{\rm MJ}$ represent the efficiency of events which pass the loosened lepton identification criteria to also pass the standard identification criteria, and are measured in control regions dominated by prompt leptons and MJ events, respectively.


\subsection{Event yields\label{ssec:yields}}
We split the selected \ljets\ events into subsamples according to lepton flavor ($e$ or $\mu$), jet multiplicity, and the number of $b$-tagged jets in the event to verify an adequate description of the data with our signal and background model. In general, we observe good agreement between data and simulations, and systematic uncertainties on the final result explicitly account for moderate agreement observed in some kinematic distributions~(cf.~Sec.~\ref{sec:syst}).

\begin{figure}
\includegraphics[width=0.235\textwidth]{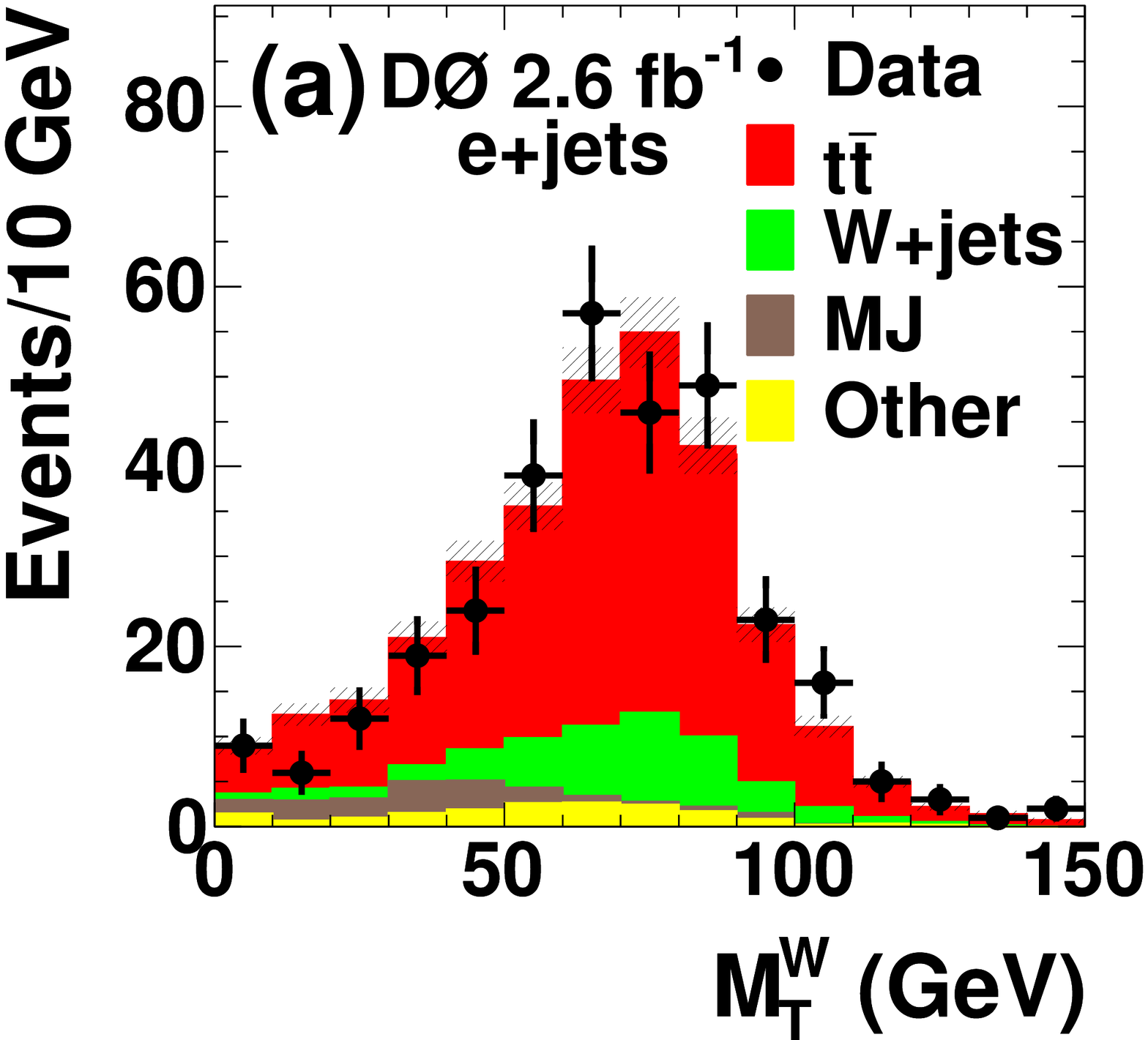}
\hspace{-1mm}
\includegraphics[width=0.235\textwidth]{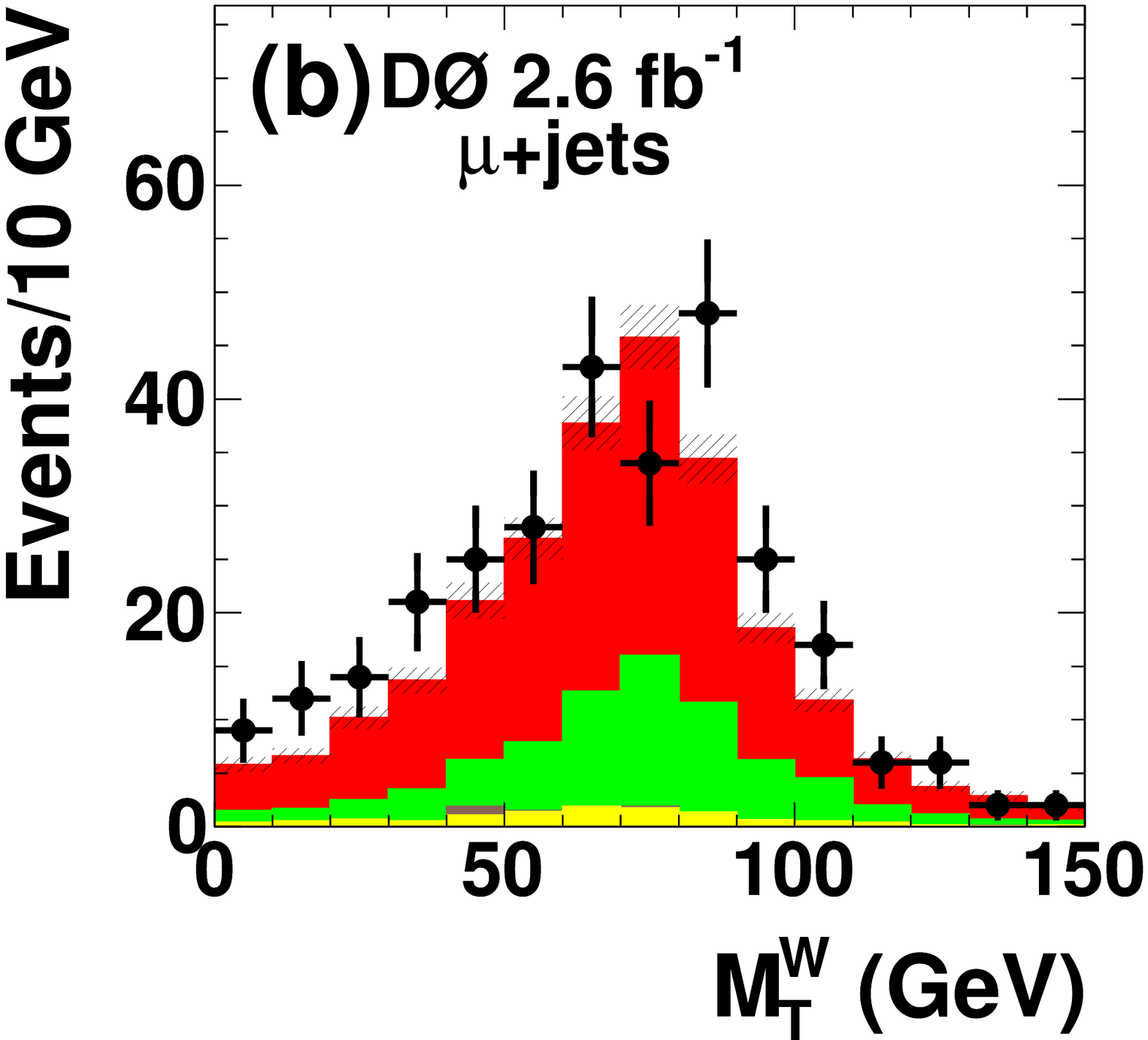}\\
\includegraphics[width=0.235\textwidth]{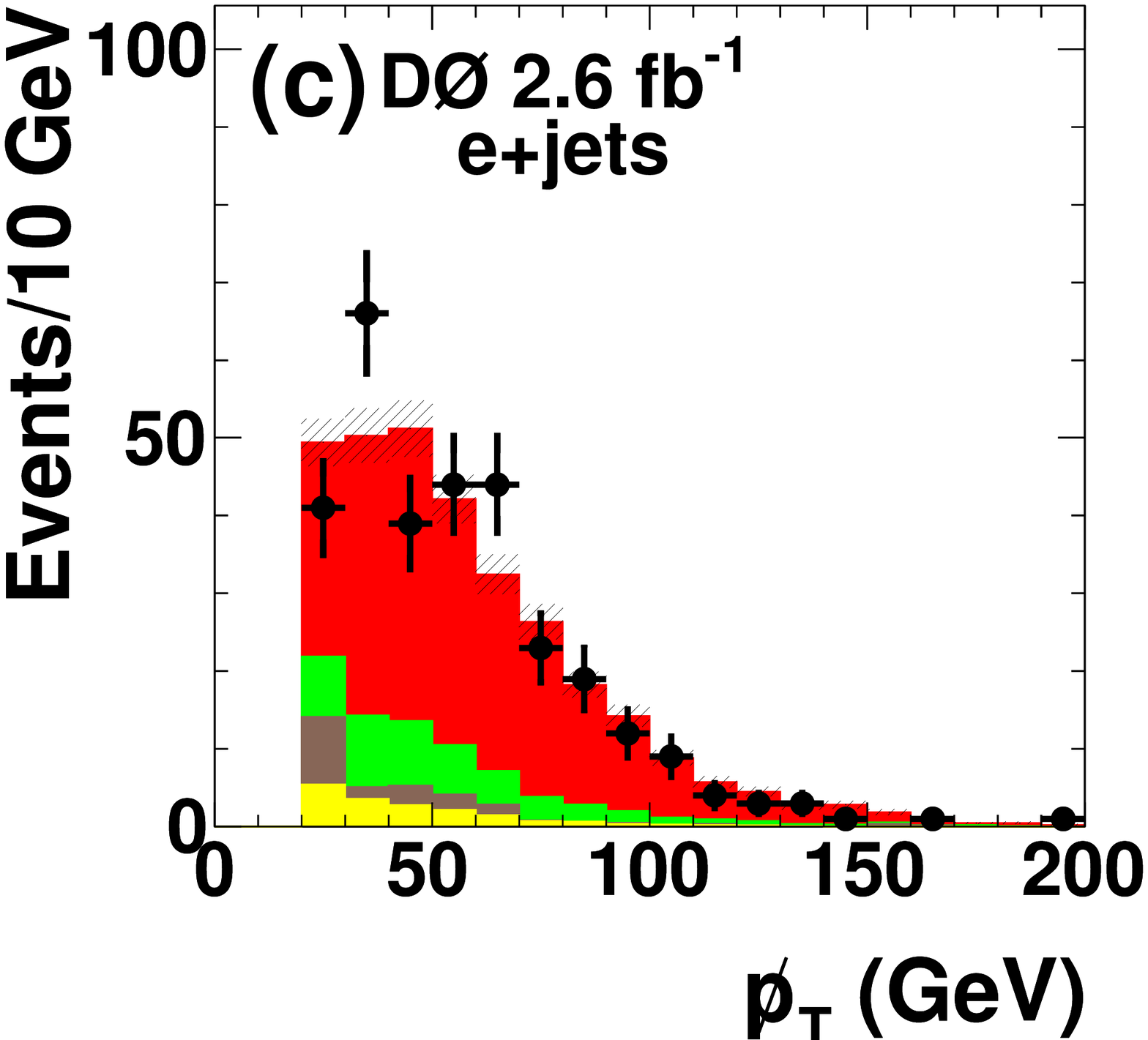}
\hspace{-1mm}
\includegraphics[width=0.235\textwidth]{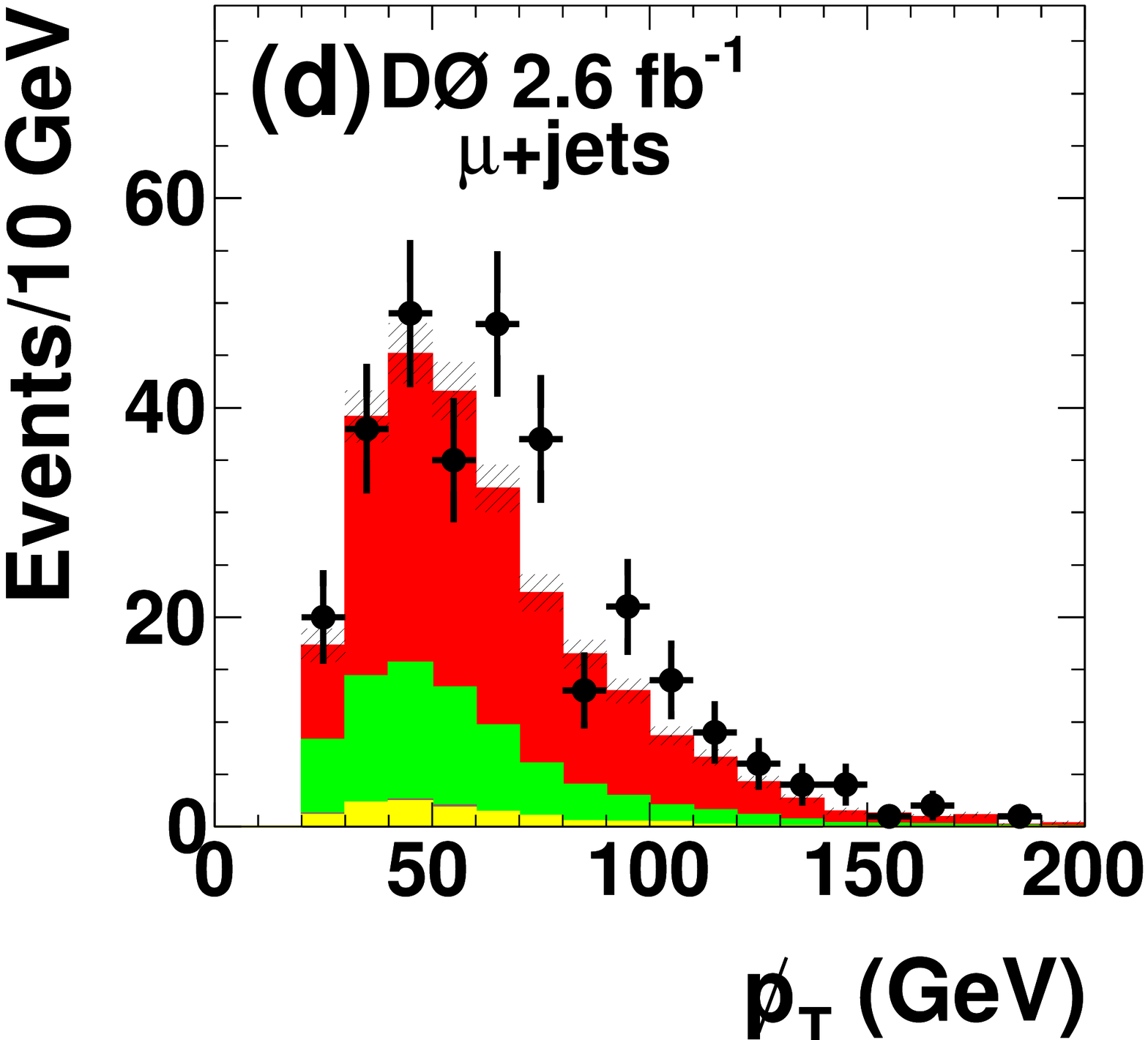}
\caption{\label{fig:yield}
The transverse mass of the  $W$ boson $M_T^W$ for events with at least one $b$-tag is shown for the (a)~\ejets and (b)~\mujets channels. 
Similarly, $\met$ is shown for the (c)~\ejets and (d)~\mujets channels. 
The statistical uncertainties on the prediction from the \ttbar signal and background models are indicated by the hatched area.
}
\end{figure}

The numbers of events surviving the final stage of selection with at least one $b$-tag are summarized in Table~\ref{tab:yields}. 
Here, for ease of comparison, the contributions from $\ttbar$ events are scaled to $7.45^{+0.5}_{-0.7}$\,pb, the NLO cross section including NNLO approximations~\cite{bib:ttxsec}. The total $\wjets$ cross section is adjusted to bring the absolute yield from our signal and background model into agreement with the number of events selected in data before applying $b$-jet identification criteria. The distributions in the transverse mass of the $W$ boson, $M_T^W$~\cite{bib:mwtrans}, and in $\met$ are shown in Fig.~\ref{fig:yield} for data with at least one $b$-tag, together with the predictions from our signal and background models. 

\begin{table}[ht]
\caption{\label{tab:yields}
Numbers of events selected in data, compared to yield predictions for individual processes using simulations, in the \ejets and \mujets channels with exactly 4 jets and at least one $b$-tagged jet, split according to \mbox{$b$-tag} multiplicity. Uncertainties are purely statistical. See text for details.
}
\begin{center}
\begin{tabular}{l|r@{ $\pm$ }r  r@{ $\pm$ }r }
\hline
      \hline
 & \multicolumn{2}{c}{1 $b$-tag} &
 \multicolumn{2}{c}{$>$1 $b$-tags } \\
\hline 
\ejets\\
\quad $t\bar{t}$   &~139.2 & 3.0&~~~~91.8 & 2.5 \\
\quad \wjets       &  39.9 & 1.2  &   4.7 & 0.3 \\
\quad MJ           &  23.5 & 2.1  &   5.7 & 1.0 \\
\quad \zjets       &   7.6 & 0.7  &   0.9 & 0.1 \\
\quad Other        &   6.6 & 0.4  &   1.9 & 0.1 \\
\quad Total        & 216.7 & 3.9  & 105.1 & 2.7 \\
\quad Observed~    & \multicolumn{2}{c}{223} & \multicolumn{2}{c}{89} \\ 
\hline
\mujets\\
\quad $t\bar{t}$   & 105.9 & 2.4 & 70.9 & 2.0 \\
\quad \wjets       &  59.9 & 1.8 &  7.2 & 0.5 \\
\quad MJ           &   5.2 & 0.9 &  2.0 & 0.6 \\
\quad \zjets       &   5.3 & 0.5 &  1.2 & 0.2 \\
\quad Other        &   5.0 & 0.3 &  1.3 & 0.1 \\
\quad Total        & 181.3 & 3.2 & 82.6 & 2.2 \\
\quad Observed     & \multicolumn{2}{c}{191} & \multicolumn{2}{c}{112} \\ 
\hline
\hline
\end{tabular}
\end{center}
%
%
\end{table}

\section{General description of the method\label{sec:method}}
In this section, we describe the measurement of \dm using the ME method. The procedure is similar to the one used in Ref.~\cite{Aba06,bib:me26fb} to measure the average top quark mass $\mtop$, but instead of simultaneously determining \mtop and the jet energy scale (JES), here we measure directly the masses of the top and antitop quarks, \mt and \mtb, which provides $\dm$ and $\msum$. 
We review the ME approach in Sec.~\ref{ssec:prob}, the calculation of signal and background event probabilities in Secs.~\ref{ssec:psig} and~\ref{ssec:pbkg}, respectively, as well as the parametrization of the detector response and the use of $b$-tagging information in Sec.~\ref{ssec:tf}.

\subsection{Probability densities for events\label{ssec:prob}}
To optimize the use of kinematic and topological information, each event is assigned a probability \pevt to observe it as a function of the assumed top and antitop quark masses: $\pevt=\pevt(\mt,\mtb)$. The individual probabilities for all events in a given sample are combined to form a likelihood, from which the \dm and \msum parameters are extracted. Simplifying assumptions are made in the expression of the likelihood about, e.g., detector response or the sample composition, are made to render the problem numerically solvable. It is therefore necessary to calibrate the method using fully simulated MC events, as detailed in Sec.~\ref{ssec:calib}. Systematic uncertainties are estimated to account for possible effects of these assumptions on the extracted value of \dm.

Assuming that the signal and background physics processes do not interfere, the contribution to the overall probability from a single event can be formulated as
\begin{eqnarray}
\pevt(x;\mt,\mtb,\fsig) &=& \acc(x)\{~\fsig\cdot\psig(x;\mt,\mtb)\nonumber\\
                        &&+~(1-\fsig)\cdot\pbkg(x)~ \}\,,\label{eq:pevt}
\end{eqnarray}
where $x$ denotes the set of measured kinematic variables for the event observed in the detector, \fsig is the fraction of signal events in the sample, $\acc(x)$ reflects the detector acceptance and efficiencies for a given $x$, and $\psig$ and $\pbkg$ are the probabilities for the event to arise from \ttbar or \wjets production, respectively. The production of $W$ bosons in association with jets is the dominant background, and we neglect all other contributions to $\pbkg$.
Kinematically similar contributions from other background processes like MJ production are accounted for in the analysis implicitly (cf.~Sec.~\ref{sec:syst}).

Both signal and background probabilities depend on the JES, which is defined as the ratio of the calibrated energy of a jet over its uncalibrated energy. The standard calibration of jet energies accounts for the energy response of the calorimeters, the energy that crosses the cone boundary due to the transverse shower size, and the additional energy from pileup of events and from multiple $p\bar p$ interactions in a single beam crossing. Although the $\dm$ observable is not expected to show a strong dependence on JES by construction, we apply an additional absolute calibration to the JES using a matrix element which is a function of \mtop and JES from Refs.~\cite{Aba06,bib:me26fb}. 
The potential systematic bias on \dm from the uncertainty on the absolute value of the JES is estimated in Sec.~\ref{sec:syst}.

To extract the masses \mt and \mtb from a set of $n$ selected events, with sets of measured kinematic quantities $x_{1},...,x_{n}$, a likelihood function is defined from the individual event probabilities according to Eq.~(\ref{eq:pevt}):
\begin{equation}
L(x_{1},...,x_{n};\, \mt,\mtb,\fsig)  =  \prod_{i=1}^{n} \pevt(x_{i};\, \mt,\mtb,\fsig)\,.\label{eq:lhmtmtb}
\end{equation}
For every assumed $(\mt,\mtb)$ pair, we first determine the value of $\fsig\equiv\fsig^{\rm best}$ that maximizes this likelihood.

\subsection{Calculation of signal probability $\boldsymbol P_{\bf sig}$\label{ssec:psig}}
The probability density for the signal to yield a given set of partonic final state four-momenta $y$ in $p\bar p$ collisions is proportional to the differential cross section $\dif\sigma$ for $\ttbar$ production:
\begin{eqnarray}
\dif\sigma&&\!\!\!\!\!\!\!\!\!\!(p\bar p\to\ttbar\to y;\mt,\mtb)
            = \int\limits_{q_{1},\,q_{2}}\sum_{{}^{\rm quark}_{\rm flavors}}\dif q_{1}\dif q_{2} f(q_{1})f(q_{2}) \nonumber\\
          &&\qquad\qquad\quad\times \frac{(2\pi)^{4}\left|\mathcal{M}(q\bar q\to \ttbar\to y)\right|^{2}}{2q_{1}q_{2}s}\dif\Phi_{6}\,,\label{eq:dsigma}
\end{eqnarray}
where $\mathcal{M}$ denotes the matrix element for the $q\bar{q}\to t\bar{t}\to b(l\nu)\bar b(q\bar q')$ process, $s$ is the square of the center-of-mass energy, $q_{i}$ is the momentum fraction of the colliding parton $i$ (assumed to be massless), and ${\rm d}\Phi_{6}$ is an infinitesimal element of six-body phase space. The $f(q_i)$ denote the probability densities for finding a parton of given flavor and momentum fraction $q_i$ in the proton or antiproton, and the sum runs over all possible flavor configurations of the colliding quark and antiquark. In our definition of $\mathcal M$, and therefore the \ttbar signal probability, only quark-antiquark annihilation at LO is taken into account; in this sense, Eq.~(\ref{eq:dsigma}) does not represent the full differential cross section for $\ttbar$ production in $p\bar p$ collisions. Effects from gluon-gluon and quark-gluon induced \ttbar production are accounted for in the calibration procedure described in Sec.~\ref{ssec:calib}. We further test for an effect on \dm from higher-order corrections in Sec.~\ref{sec:cross}.

The differential cross section for observing a \ttbar event with a set of kinematic quantities $x$ measured in the detector can be written as
\begin{eqnarray}
&&\!\!\!\!\!\!\!\!\!\!\!\!\!\!\dif\sigma(p\bar p\to\ttbar\to x;\mt,\mtb,\jes)\nonumber\\
          &&\!\!\!=\acc(x)\!\!\int_{y}\!\!\dif y\,\dif\sigma(p\bar p\to\ttbar\to y;\mt,\mtb)W(x,y;\jes)\,,
\end{eqnarray}
where finite detector resolution and offline selections are taken explicitly into account through the convolution over a transfer function $W(x,y;\jes)$ that defines the probability for a partonic final state $y$ to appear as $x$ in the detector given an absolute JES correction \jes.

With the above defintions, the differential probability to observe a $\ttbar$ event with a set of kinematic quantities $x$ measured in the detector is given by
\begin{eqnarray}
\!\!\!\!\!\!\!\!\psig(x;\mt,\mtb,\jes) &=& \frac{\dif\sigma(p\bar p\to\ttbar\to x;\mt,\mtb,\jes)}
                                {\sigma_{\rm obs}(p\bar p\to\ttbar;\mt,\mtb,\jes)}\,,\label{eq:psig}
\end{eqnarray}
where $\sigma_{\rm obs}$ is the cross section for observing $\ttbar$ events in the detector for the specific ME $\mathcal M$ defined in Eq.~(\ref{eq:dsigma}):
\begin{eqnarray}
    &&\!\!\!\!\!\!\!\!\!\!\!\!\!\sigma_{\rm obs}(p\bar p\to\ttbar;\mt,\mtb,\jes) \nonumber\\
    &&\!\!\!\! = \int_{x,y}\!\!\!\!\!\!{\dif x\,\dif y~\dif\sigma(p\bar p\to\ttbar\to y;\mt,\mtb)}W(x,y;\jes)\acc(x)\,\nonumber\\
    &&\!\!\!\! = \int_{y}\!\!{\dif y~\dif\sigma(p\bar p\to\ttbar\to y;\mt,\mtb)}\!\!\int_{x}\!\!{\dif x~W(x,y;\jes)\acc(x)}\,.\nonumber
\end{eqnarray}
%
The normalization factor $\sigma_{\rm obs}$ is calculated using MC integration techniques:
\begin{eqnarray}
\!\!\!\!\!\!\!\!\!\!\!\!\sigma_{\rm obs}(p\bar p\to\ttbar;\mt,\mtb,\jes) 
                    \!\!&\simeq&\!\! \sigma_{\rm tot}(\mt,\mtb)\!\times\!\langle\acc|\mt,\mtb\rangle,\label{eq:accapprox}
\end{eqnarray}
where
\begin{equation}
\sigma_{\rm tot}(\mt,\mtb) = \int_{y}\!\!{\dif y~\dif\sigma(p\bar p\to\ttbar\to y;\mt,\mtb)}\,, \label{eq:acc}\nonumber
\end{equation}
and
\begin{equation}
\langle\acc|\mt,\mtb\rangle\equiv\frac1{N_{\rm gen}}\sum_{\rm acc}\omega\,. \label{eq:acccalc}\nonumber
\end{equation}

To calculate the $\langle\acc|\mt,\mtb\rangle$ term, events are generated according to $\dif\sigma(p\bar p\to\ttbar;\mt,\mtb)$ using {\sc pythia} and passed through the full simulation of the detector. Here, $N_{\rm gen}$ is the total number of generated events, $\omega$ are the MC event weights that account for trigger and identification efficiencies, and the sum runs over all accepted events.

The formulae used to calculate the total cross section $\sigma_{\rm tot}$ and the matrix element $\mathcal M$ are described below in Secs.~\ref{sssec:xsec} and~\ref{sssec:me}. In all other respects, the calculation of the signal probability proceeds identically to that in Refs.~\cite{Aba06,bib:me26fb}, with the following exceptions: $(i)$~CTEQ6L1 PDFs are used throughout, and
$(ii)$~the event probabilities are calculated on a grid in \mt and \mtb spaced at 1~GeV intervals along each axis. As described in Sec.~\ref{ssec:massfit}, a transformation of variables to $\dm$ and \msum is performed when defining the likelihood. 

\subsubsection{Calculation of the total cross section $\sigma_{\rm tot}$\label{sssec:xsec}}
Without the assumption of equal top and antitop quark masses, the total LO cross section for the $q\bar q\to\ttbar$ process in the center of mass frame is given by
\begin{equation}\label{eq:sigmatot}
\sigma=\frac{16\pi\alpha_{s}^{2}}{27s^{\frac{5}{2}}}|\vec p|\left[3E_{t}E_{\bar t}+|\vec p|^2+3\mt\mtb\right],
\end{equation}
where $E_{t}$~($E_{\bar t}$) are the energies of the top and antitop quark, and $\vec p$ is the three-momentum of the top quark. This reduces to the familiar form for $\mt=\mtb$:
\[
\sigma=\frac{4\pi\alpha_{s}^{2}}{9s}\beta\left(1-\frac{\beta^{2}}{3}\right),
\]
where $\beta=|\vec p_{t}|/E_{t}=|\vec p_{\bar t}|/E_{\bar t}$ represents the velocity of the $t$ (or $\bar{t}$) quark in the $q\bar q$ rest frame.

\begin{figure}
\begin{centering}
\includegraphics[width=0.49\textwidth]{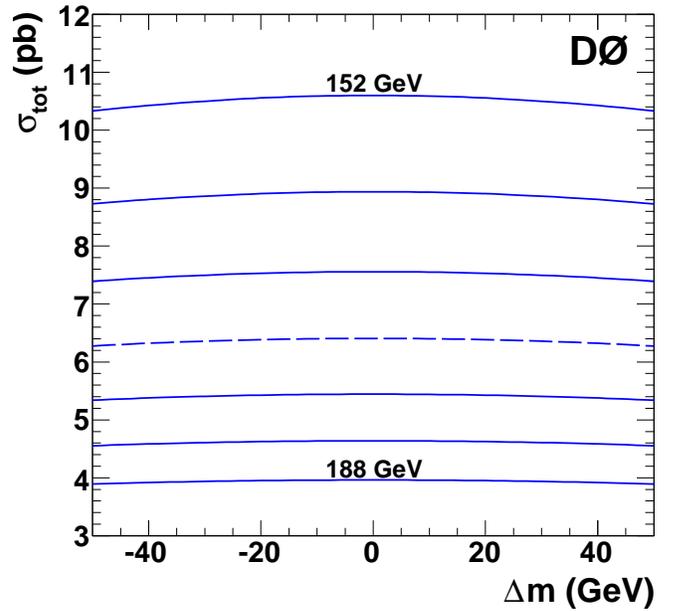}
\par\end{centering}
\caption{\label{fig:xsec}
The total $p\bar{p}\to t\bar{t}$ production cross section $\sigma_{\rm tot}$ defined in Eq.~(\ref{eq:acc}) as a function of $\dm$ and $\msum$. 
Each line shows $\sigma_{\rm tot}$ as a function of $\dm$ for a given value of $\msum$ displayed above the curve. The range from $152$~GeV to $188$~GeV is shown in $6$~GeV increments, the broken line corresponds to 170~GeV.
}
\end{figure}

\begin{figure}
\begin{centering}
\includegraphics[width=0.24\textwidth,clip=]{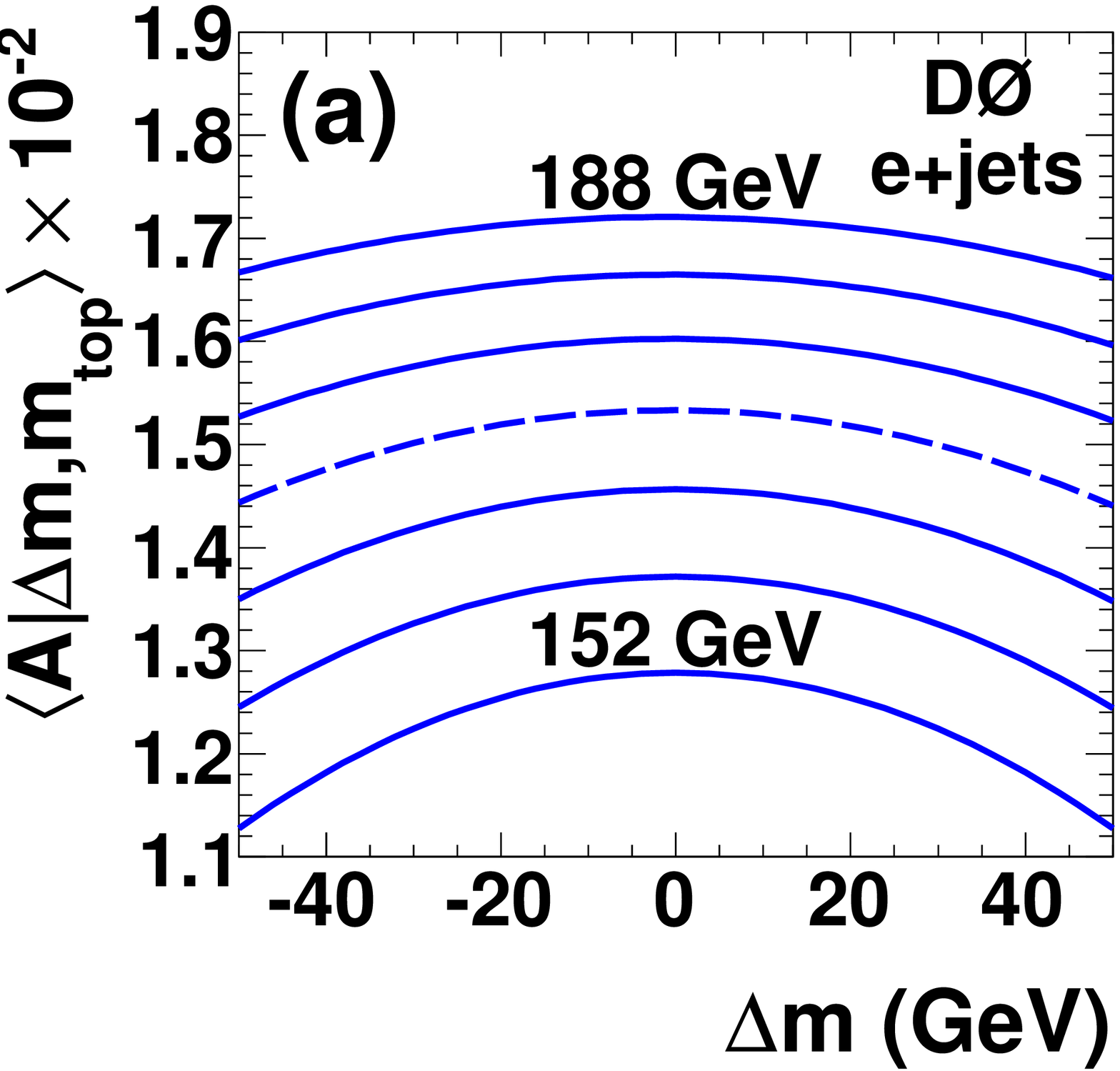}
\hspace{-1.5mm}
\includegraphics[width=0.24\textwidth,clip=]{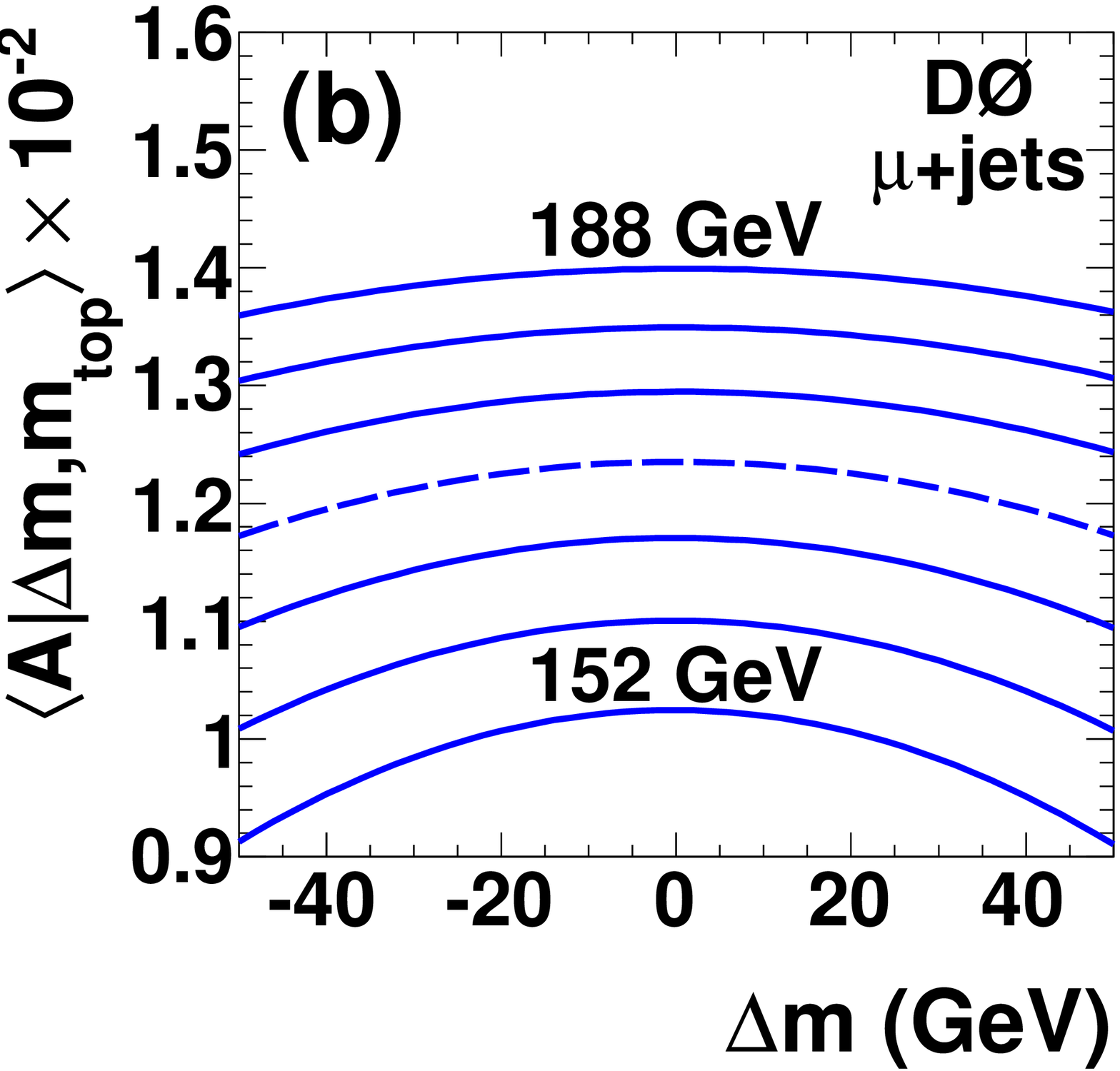}
\par\end{centering}
\caption{\label{fig:acc}
The dependence of the overall average acceptance $\langle\acc|\mt,\mtb\rangle$ on $\dm$ and $\msum$, as defined in Eq.~(\ref{eq:acccalc}), for the (a)~\ejets and (b)~\mujets signal MC samples. Each line shows $\langle\acc|\mt,\mtb\rangle$ as a function of $\dm$ for a given value of $\msum$ displayed above the curve. The range from $152$~GeV to $188$~GeV is shown in $6$~GeV increments, the broken lines correspond to 170~GeV.
}
\end{figure}

Integrating Eq.~(\ref{eq:sigmatot}) over all incoming $q\bar{q}$ momenta and using the appropriate PDF yields $\sigma_{\rm tot}(p\bar{p}\to t\bar{t};\,\mt,\mtb)$, as defined for any values of \mt and \mtb in Eq.~(\ref{eq:acc}). Figure~\ref{fig:xsec} displays the dependence of $\sigma_{\rm tot}$ on \dm for a given \msum. The corresponding average acceptance term $\langle\acc|\mt,\mtb\rangle$, as defined in the same equation, is shown in Fig.~(\ref{fig:acc}) for the \ejets and \mujets channels.


\subsubsection{Calculation of the matrix element $\mathcal M$\label{sssec:me}}
The LO matrix element for the $q\bar q\to\ttbar$ process we use in our analysis is
\begin{eqnarray}
\left|\mathcal{M}\right|^{2} &=& \frac{g_{s}^{4}}{9}F\bar{F}\cdot\frac{2}{s}\nonumber\\
                             &\times&\!\!\!\!\!\left\{(E_{t}-|\vec p_{t}|c_{qt})^{2}+(E_{\bar t}+|\vec p_{\bar t}|c_{qt})^{2}+2\mt \mtb\right\}.
\label{eq:me}
\end{eqnarray}
The form factors $F\bar F$ are identical to those given in Eqs.~(24) and~(25) of Ref.~\cite{Aba06}. For the special case of $\mt=\mtb$, the expression in Eq.~(\ref{eq:me}) reduces to
\[
\left|\mathcal{M}\right|^{2}=\frac{g_{s}^{4}}{9}F\bar{F}\cdot\left(2-\beta^{2}s_{qt}^{2}\right),
\]
which is identical to Refs.~\cite{Aba06,bib:mecalc}, where $s_{qt}$ is the sine of the angle between the incoming parton and the outgoing top quark in the $q\bar q$ rest frame.

\subsection{Calculation of the background probability $\boldsymbol P_{\bf bkg}$\label{ssec:pbkg}}
The expression for the background probability $\pbkg$ is similar to that for $\psig$ in Eq.~(\ref{eq:psig}), except that the ME $\mathcal{M}_{\wjets}$ is for \wjets production, and all jets are assumed to be light quark or gluon jets. Clearly, $\mathcal{M}_{\wjets}$ does not depend on $\mt$ or $\mtb$, and $\pbkg$ is therefore independent of either. We use a LO parameterization of $\mathcal{M}$ from the \vecbos~\cite{bib:VECBOS} program. More details on the calculation of the background probability can be found in Ref.~\cite{Aba06}.

\subsection{Description of detector response\label{ssec:tf}}
The transfer function $W(x,y,\jes)$, which relates the set of variables $x$ characterizing the reconstructed final-state objects to their partonic quantities $y$, is crucial for the calculation of the signal probability according to Eq.~(\ref{eq:psig}), and the corresponding expression for \pbkg. A full simulation of the detector would not be feasible for calculating event probabilities because of the overwhelming requirements for computing resources. Therefore, we parametrize the detector response and resolution through a transfer function. 

In constructing the transfer function, we assume that the functions for individual final-state particles are not correlated. We therefore factorize the transfer function into contributions from each measured final-state object used in calculating \psig, that is the isolated lepton and four jets. The poorly measured imbalance in transverse momentum \met, and consequently the transverse momentum of the neutrino, is not used in defining event probabilities. We assume that the directions of $e$, $\mu$, and jets in $(\eta,\phi)$ space are well-measured, and therefore define the transfer functions for these quantities as $\delta$ functions: $\delta^2(\eta,\phi)\equiv\delta(\eta_y-\eta_x)\delta(\phi_y-\phi_x)$. This reduces the number of integrations over the 6-particle phase space $\dif\Phi_6$ by $5\times2=10$ dimensions. The magnitudes of particle momenta $|\vec p|$ display significant variations in resolution for leptons and jets and are therefore parameterized by their corresponding resolutions.

There is an inherent ambiguity in assigning jets reconstructed in the detector to specific partons from $\ttbar$ decay. Consequently, all 24 permutations of jet-quark assignments are considered in the analysis. The inclusion of $b$-tagging information provides improved identification of the correct permutation.  This additional information enters the probability calculation through a weight $w_i$ on a given permutation $i$ of jet-parton assignments. 
The $w_i$ are larger for those permutations that assign the $b$-tagged jets to $b$ quarks and untagged jets to light quarks. The sum of weights is normalized to unity: $\sum_{i=1}^{24}\!w_i = 1$.

Based on the above, we define the transfer function as
\begin{eqnarray}
  \label{eq:tfdefinition-btag}
    &&\!\!\!\!\!\!\!\!\!\!\!\!\!\!\!\!\!\!\!\!\!\!\!\!\!\!
    W(x,y;\,\jes) = W_{\ell}(E_x,E_y)\delta_\ell^2(\eta,\phi) \nonumber\\
    &&\!\!\!\!\!\!\!\!\!\!\!\!\!\!\!\!
             \times \sum_{i=1}^{24}\!w_i 
             \left\{
               \prod_{j=1}^{4}\,\delta_{ij}^2(\eta,\phi) W_{\rm jet}(E^i_x,E^j_y;\jes)
             \!\!\right\}\!,
\end{eqnarray}
where $\ell$ denotes the lepton flavor, with a term $W_e$ describing the energy resolution for electrons and $W_\mu$ the resolution in the transverse momentum for muons. Similarly, $W_{\rm jet}$ describes the energy resolution for jets. The sum in $i$ is taken over the 24 possible permutations of assigning jets to quarks in a given event. More details on $W_{\ell}$ and $W_{\rm jet}$ can be found in Ref.~\cite{bib:me26fb}.

The weight $w_i$ for a given permutation $i$ is defined by a product of individual weights $w_i^j$ for each jet $j$. 
For $b$-tag\-ged jets, $w_i^j$ is equal to the per-jet  tagging efficiency $\epsilon_{\rm tag}(\alpha_k;\ \etj,\,\eta^j)$, where $\alpha_k$ labels the three possible parton-flavor assignments of the jet: $(i)$~$b$~quark, $(ii)$~$c$~quark, and $(iii)$~light ($u,d,s$) quark or gluon.    For untagged jets, the $w_i^j$ factors are equal to  $1-\epsilon_{\rm tag}(\alpha_k;\ \etj,\,\eta^j)$.



Because the contributions to \wjets are parameterized by $\mathcal{M}_{\wjets}$ without regard to heavy-flavor content, the weights $w_i$ for each permutation in the background probability are all set equal. 

\section{Measurement of the top-antitop quark mass difference\label{sec:measurement}}
\subsection{Fit to the top-antitop quark mass difference\label{ssec:massfit}}
For the set of selected events, the likelihood $L(\mt,\mtb)$ is calculated from Eq.~(\ref{eq:lhmtmtb}) (Sec.~\ref{ssec:prob}). The signal fraction $\fsig^{\rm best}$ that maximizes the likelihood is determined at each $(\mt,\mtb)$ point for grid spacings of 1~\GeV. Subsequently, a transformation is made to the more appropriate set of variables $(\dm,\msum)$:
\begin{eqnarray}
L(x_{1},&&\!\!\!\!\!\!\!\!\!\!\!\!\!\!\!\!...,x_{n};\dm,\msum) \nonumber\\
&=& L[x_{1},...,x_{n};\,\dm,\msum,\fsig^{\rm best}(\dm,\msum)]\,.\label{eq:lh}
\end{eqnarray}

To obtain the best estimate of $\dm$ in data, the two-dimensional likelihood in Eq.~(\ref{eq:lh}) is projected onto the $\dm$ axis, and the mean value $\langle{\dm}\rangle$, that maximizes it, as well as the uncertainty $\ddm$ on $\langle\dm\rangle$ are calculated.
%
%
%
%
This procedure accounts for any correlations between \dm and \msum.
%
As a consistency check, we simultaneously extract the average mass $\msum$ by exchanging $\dm\leftrightarrow\msum$ above.
%

\subsection{Calibration of the method\label{ssec:calib}}
We calibrate the ME method by performing 1000 MC pseudo-experiments at each input point $(\mt,\mtb)$. These are used to correlate the fitted parameters with their true input values and to assure the correctness of the estimated uncertainties. 
Each pseudo-experiment is formed by drawing $N_{\rm sig}$ signal and $N_{\rm bkg}$ background events from a large pool of fully simulated $\ttbar$ and \wjets MC events. We assume that \wjets events also represent the kinematic distributions expected from MJ production and other background processes with smaller contributions, and evaluate a systematic uncertainty from this assumption. Events are drawn randomly and can be used more than once, and an ``oversampling'' correction~\cite{bib:bootstrap} is applied. The size of each pseudo-experiment, $N=N_{\rm sig}+N_{\rm bkg}$, is fixed by the total number of events observed in the data, i.e.,~$N=312$ (303) events for the \ejets (\mujets) channel. The fraction of signal events is allowed to fluctuate relative to the signal fraction \fsig determined from data (Sec.~\ref{sub:sfrac}), assuming binomial statistics. 
The same \wjets background sample is used to form pseudo-experiments for each $(\mt,\mtb)$ mass point.

\subsubsection{Determining the signal fraction in data\label{sub:sfrac}}
The signal fraction \fsig is determined independently for the \ejets and \mujets channels directly from the selected data sample.
The likelihood depends explicitly on three parameters: \dm, \msum, and $\fsig$, as defined in Eq.~(\ref{eq:lh}). The uncalibrated signal fraction $\fsig^{\rm uncal}$ is calculated in data as an average of $\fsig^{\rm best}$ determined at each point in the $(\mt,\mtb)$ grid and weighted by the value of the likelihood at that point.
To calibrate $\fsig^{\rm uncal}$, we form 1000 pseudo-experiments for each input signal fraction $\fsig^{\rm true}$ in the interval $[0,1]$ in increments of 0.1, and extract $\fsig^{\rm uncal}$ for each one, following the same procedure as in data. Signal MC events with $\mt=\mtb=172.5~\GeV$ are used for this calibration. A linear dependence is observed between $\fsig^{\rm extr}$ and $\fsig^{\rm true}$, where $\fsig^{\rm extr}$ is the average of $\fsig^{\rm uncal}$ values extracted in 1000 pseudo-experiments for a given $\fsig^{\rm true}$. We use the results of a linear fit of $\fsig^{\rm extr}$ to $\fsig^{\rm true}$  to calibrate the fraction of signal events in data. The results are summarized in Table~\ref{tab:sfrac}. Possible systematic biases on the measured value of $\dm$ from the uncertainty on $\fsig$ are discussed in Sec.~\ref{sec:syst}.
%
%
\begin{table}[h]
\caption{\label{tab:sfrac}
Signal fractions determined from data for the assumption that $\mt=\mtb=172.5~\GeV$. The uncertainties are statistical only.
}
\begin{centering}
\begin{tabular}{rc}
\hline\hline 
Channel~~~&~Measured signal fraction \\
\hline 
\ejets~~~& $0.71~\pm~0.05$ \\
\mujets~~~& $0.75~\pm~0.04$ \\
\hline\hline
\end{tabular}
\end{centering}
\end{table}

\subsubsection{Calibration of $\dm$\label{sssec:calibdm}}
\begin{figure}[b]
\includegraphics[width=0.24\textwidth]{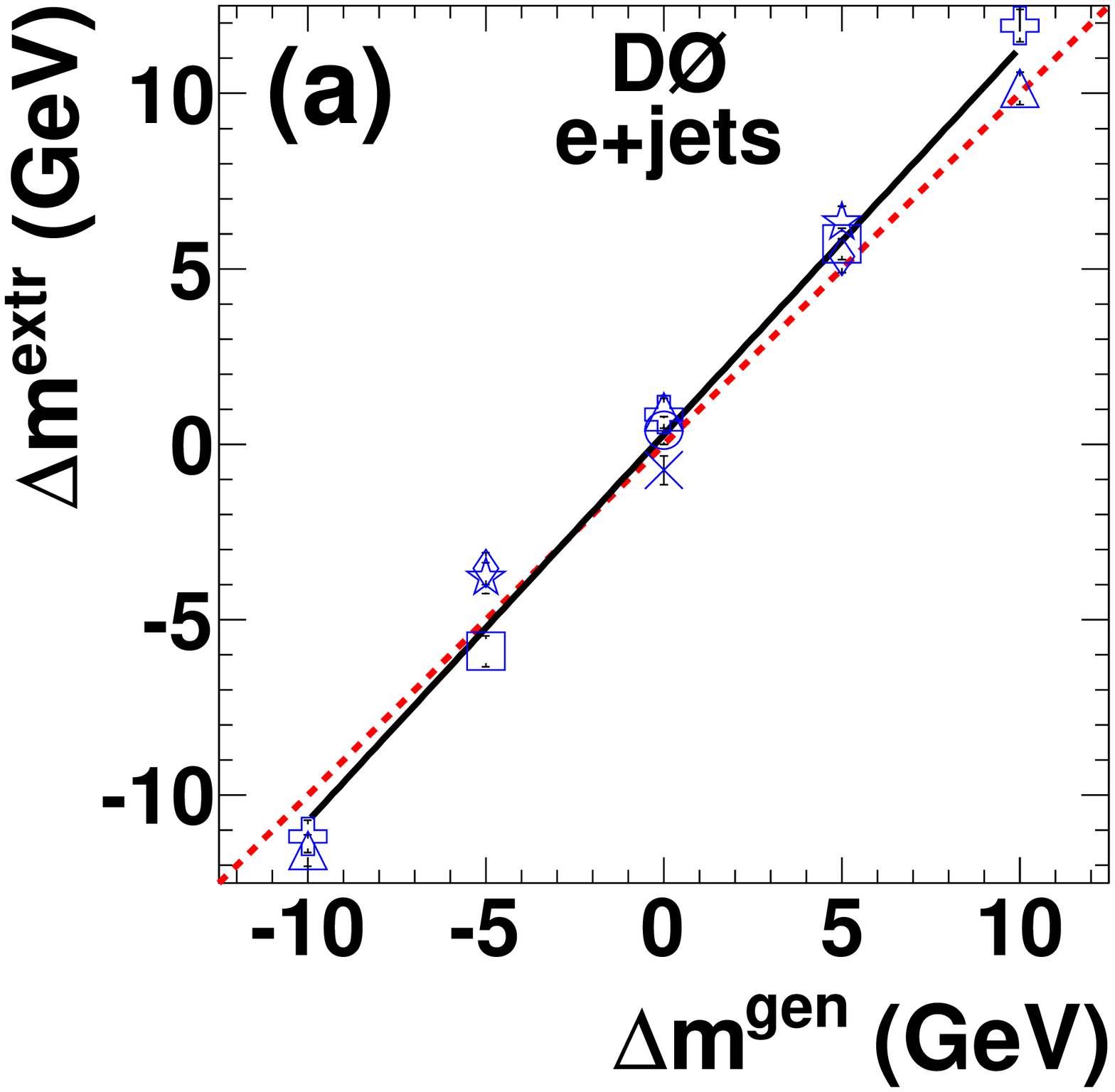}
\hspace{-2mm}
\includegraphics[width=0.24\textwidth]{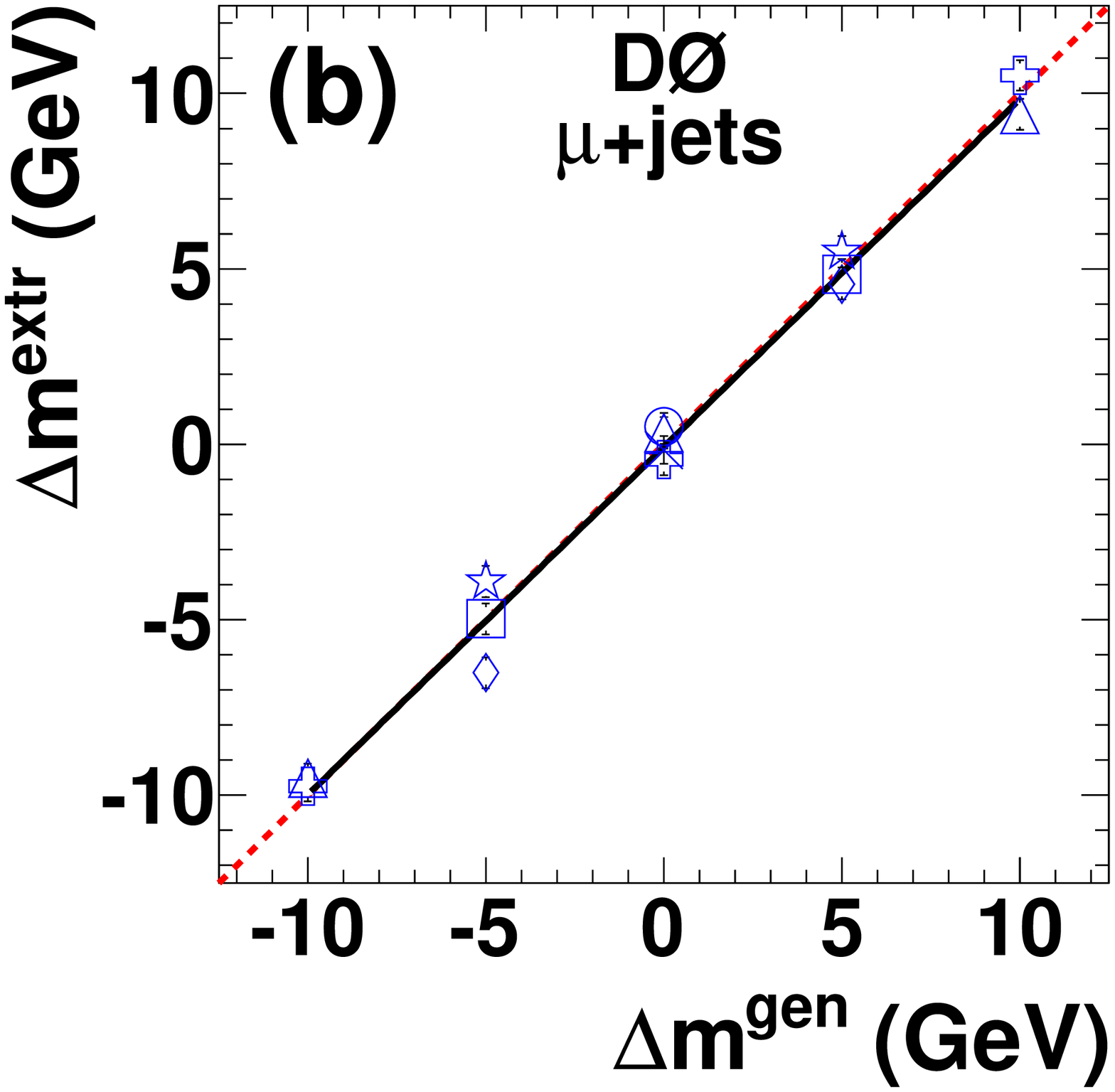}\\
\includegraphics[width=0.24\textwidth]{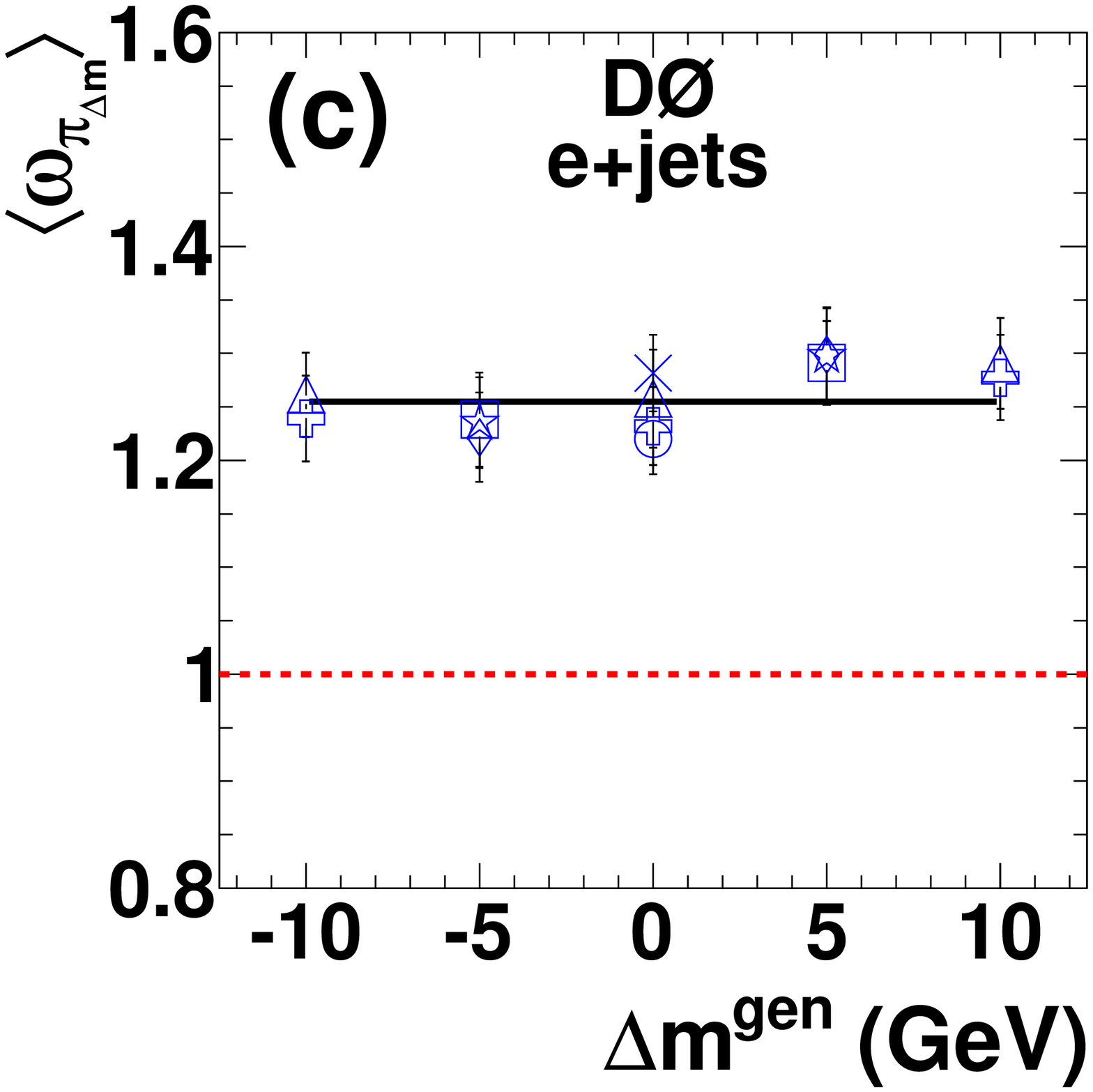}
\hspace{-2mm}
\includegraphics[width=0.24\textwidth]{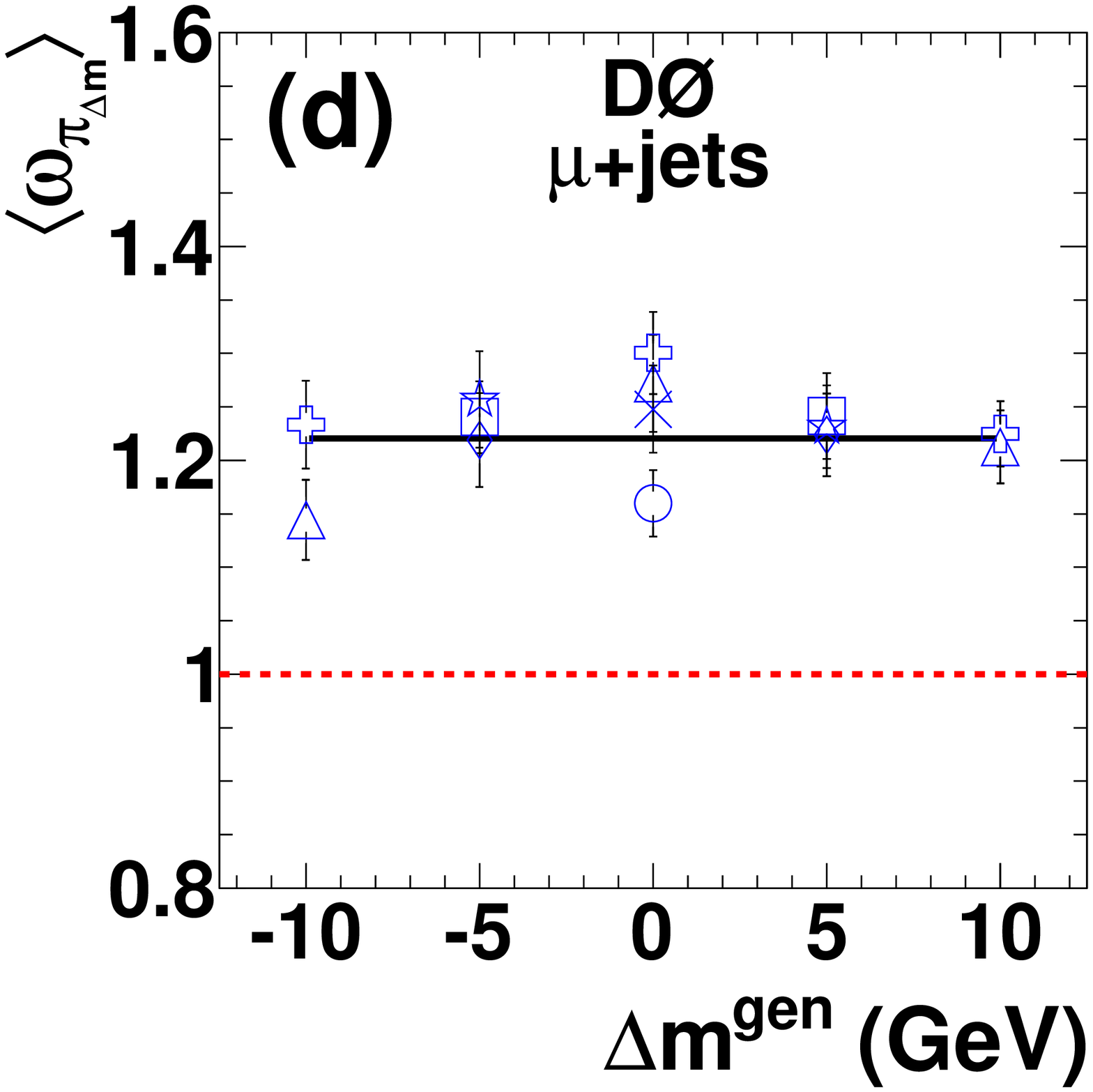}
\caption{\label{fig:calibmdel}
The calibration of the extracted $\dm$ value as a function of generated $\dm$ is shown for the (a)~\ejets and (b)~\mujets channels. The points are fitted to a linear function. Each point represents a set of 1000 pseudo-experiments for one of the fourteen $(\mt,\mtb)$ combinations. The circle, square, triangle, rhombus, cross, star, and ``$\times$'' symbols stand for $\msum=165,167.5,170,172.5,175,177.5,{\rm~and~}180~\GeV$, respectively.
Similarly, the pull widths, as defined in the text, are given for the (c)~\ejets and (d)~\mujets channels.}
\end{figure}

The dependence of the extracted \dm 
on the generated \dm is determined from the extracted values ${\dm}^{\rm extr}(\mt,\mtb)$, again obtained from averaging $\langle\dm\rangle$ over 1000 pseudo-experiments for each $(\mt,\mtb)$ combination. The resulting distribution and fit to the 14~$(\mt,\mtb)$ points is shown in Fig.~\ref{fig:calibmdel} (a) and (b) for the \ejets and \mujets channels, respectively. This provides the calibration of the extracted \dm value:
\begin{equation}\label{eq:calib}
\dm^{\rm extr}=\xi_0^{\dm}+\xi_1^{\dm}\cdot\dm^{\rm gen}\,.
\end{equation}
The fit parameters $\xi_i^{\dm}$ are summarized in Table~\ref{tab:calib}.


For an unbiased estimate of \dm and of the uncertainty \ddm on the measured $\langle\dm\rangle$ value, the distribution of the pulls should be described by a Gaussian function with a standard deviation~(SD) of unity, and centered at zero. A~SD~of the pulls larger than unity would indicate an underestimation of \ddm, which could be caused by the simplifying assumptions of the ME technique discussed in Sec.~\ref{sec:method}. For a given pseudo-experiment at $(\mt,\mtb)$, we define the pull in \dm as
\begin{equation}\label{eq:pull}
\pi_{\dm} = \frac{\langle\dm\rangle - {\dm}^{\rm extr}(\mt,\mtb)}{\ddm}\,.
\end{equation}
The pull widths $\pullw{\dm}$, defined by the SD~in Gaussian fits to the pull distributions, are also shown  for all 14 $(\mt,\mtb)$ points in Fig.~\ref{fig:calibmdel}~(c) and~(d)  for the \ejets and \mujets channels, respectively. The average pull widths $\langle\pullw{\dm}\rangle$ are taken from fits of the 14 pull widths in each channel to constant offsets and are summarized in Table~\ref{tab:calib}. We calibrate the estimated uncertainty according to $\ddm^{\rm cal}\equiv\langle\pullw{\dm}\rangle\times\ddm$.
%
%
\begin{table}
\caption{\label{tab:calib}
Fit parameters for the calibration of \dm and \msum, defined by Eq.~(\ref{eq:calib}), and average pull-widths $\langle\pullw{}\rangle$ for pulls in \dm and \msum, defined in Eq.~(\ref{eq:pull}).
}
\begin{center}
\begin{tabular}{llr@{ $\pm$ }r  r@{ $\pm$ }r  r@{ $\pm$ }r }
\hline
\hline
& Channel~ & \multicolumn{2}{c}{~~$\xi_0$~(GeV)} 
                                   & \multicolumn{2}{c}{$\xi_1$} 
                                                & \multicolumn{2}{c}{$\langle\pullw{}\rangle$} \\
\hline
\multirow{2}{*}{\dm}
          & \ejets   &~$ 0.28$ & 0.14~&~1.10 & 0.02~&~1.25 & 0.01 \\
          & \mujets  & $-0.08$ & 0.13 & 0.99 & 0.02 & 1.22 & 0.01 \\
\multirow{2}{*}{\msum}
          & \ejets   &~$ 0.53$ & 0.08~&~0.99 & 0.02~&~1.17 & 0.01 \\
          & \mujets  & $ 0.24$ & 0.07 & 1.02 & 0.02 & 1.16 & 0.01 \\
\hline
\hline
\end{tabular}
\end{center}
\end{table}

\subsubsection{Calibration of $\msum$\label{sssec:calibmavg}}
\begin{figure}[h]
\includegraphics[width=0.24\textwidth]{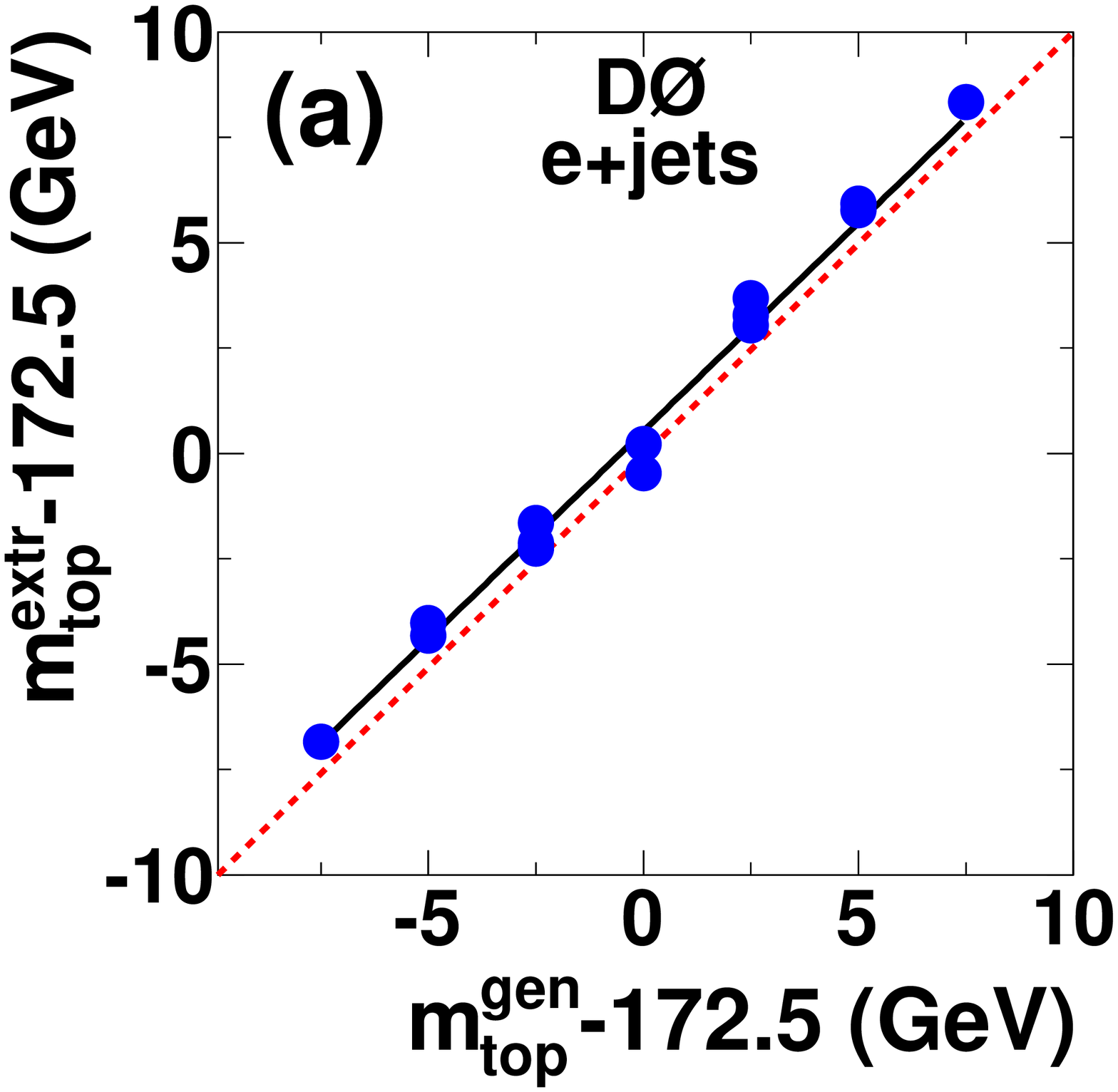}
\hspace{-2mm}
\includegraphics[width=0.24\textwidth]{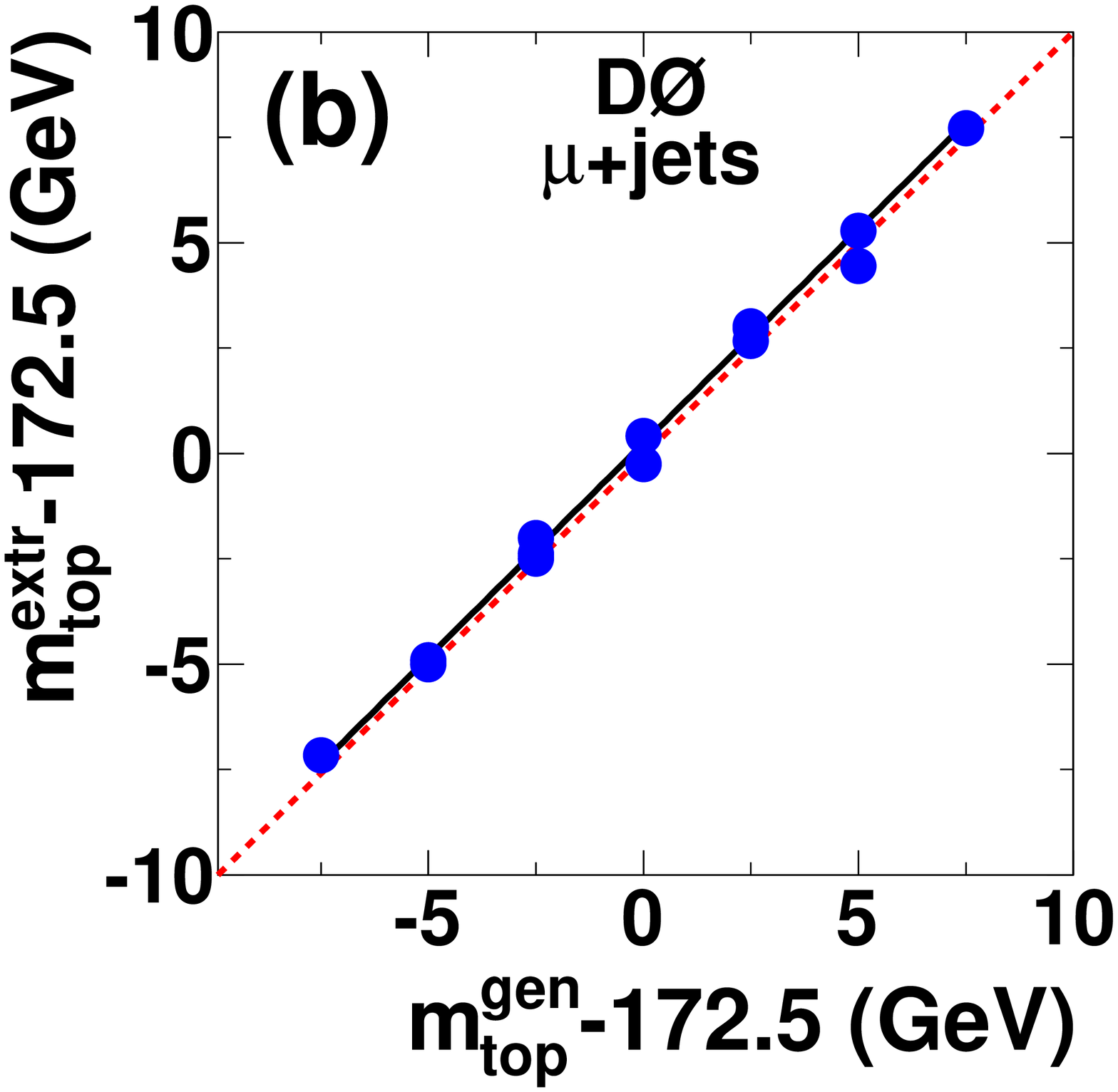}\\
\includegraphics[width=0.24\textwidth]{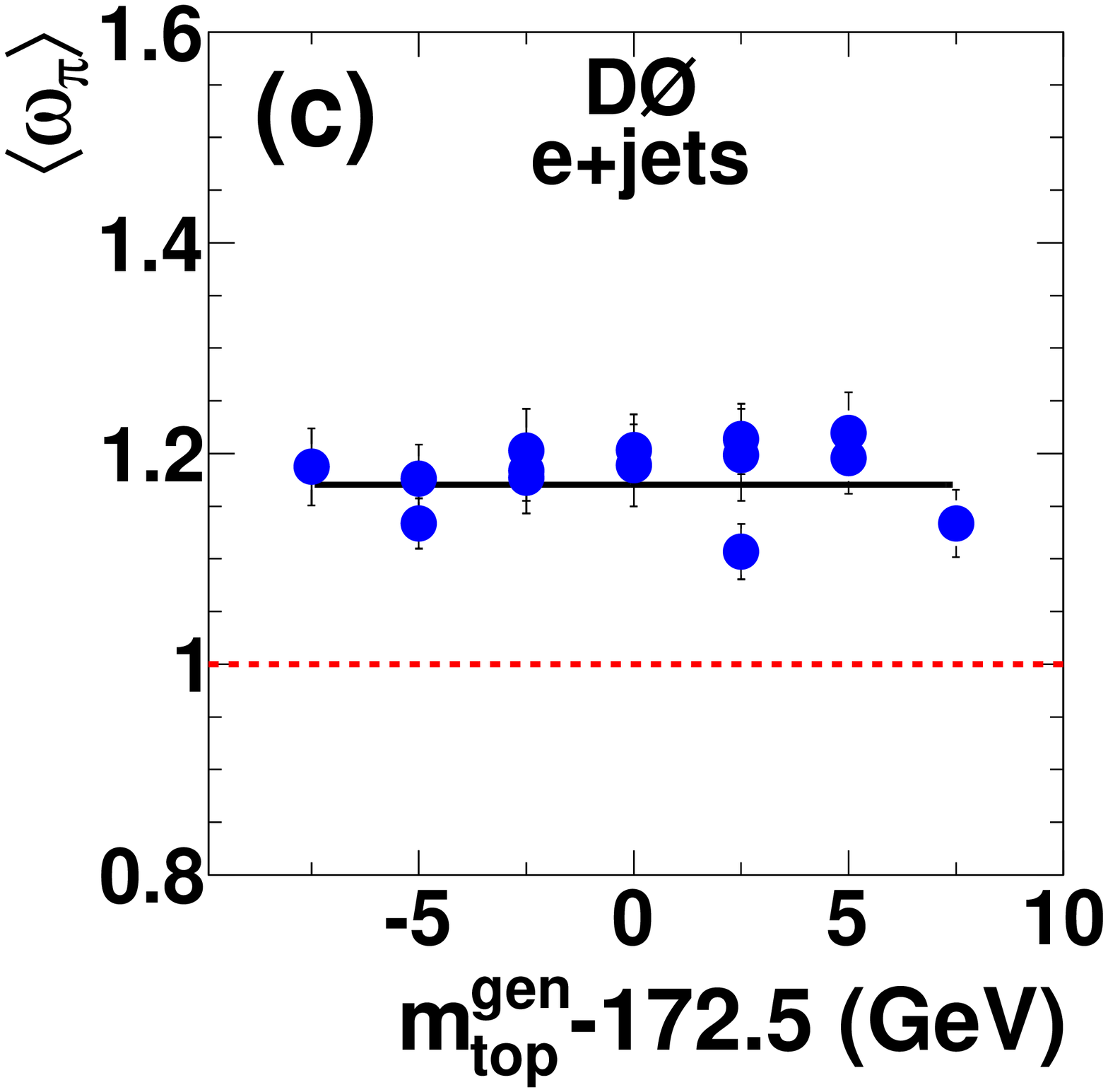}
\hspace{-2mm}
\includegraphics[width=0.24\textwidth]{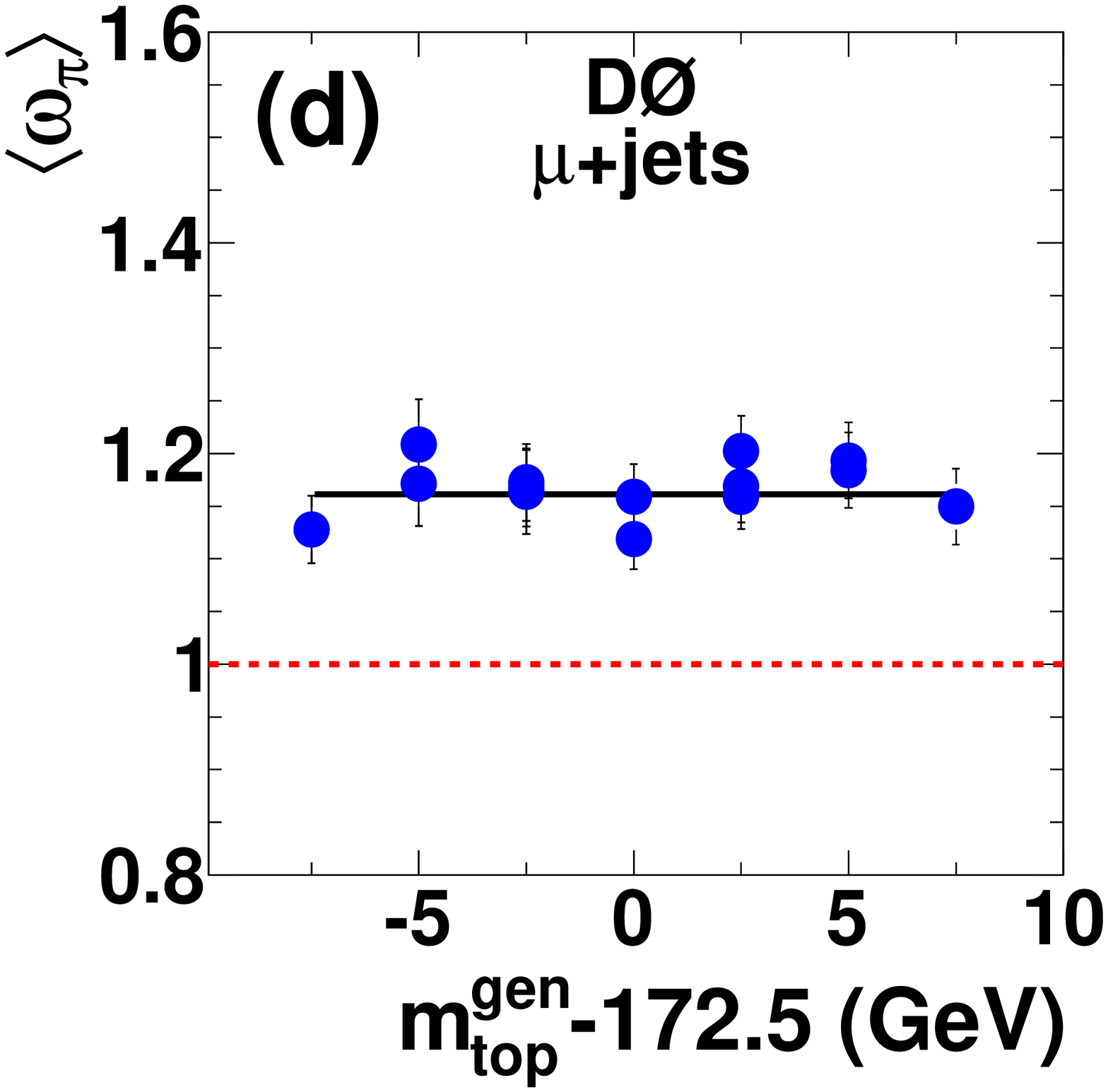}
\caption{\label{fig:calibmsum}
The calibration of the extracted $\msum$ value as a function of generated $\msum$ is shown for the (a)~\ejets and (b)~\mujets channels. The dependence is fitted to a linear function. Each point represents a set of 1000 pseudo-experiments for one of the fourteen $(\mt,\mtb)$ combinations.  
Similarly, the pull widths, as defined in the text, are given for the (c)~\ejets and (d)~\mujets channels.
}
\end{figure}

Results from an analogous calibration of $\msum$ are displayed in Fig.~\ref{fig:calibmsum}~(a) and~(b) for the \ejets and \mujets channel, respectively. The distributions in pull widths are given in parts (c) and~(d) of the same figure. The corresponding fit parameters and average pull widths are also summarized in Table~\ref{tab:calib}.

\subsection{Results\label{ssec:results}}
With the calibration of \dm and \msum, we proceed to extract \dm and, as a cross check, \msum, from the data, as described in Sec.~\ref{sec:method}. 
%
As indicated previously, the probabilities for the selected events are calculated using the ME method, and the likelihoods in \dm and \msum are constructed independently for the \ejets and \mujets channels. 

\begin{figure}
\includegraphics[width=0.24\textwidth]{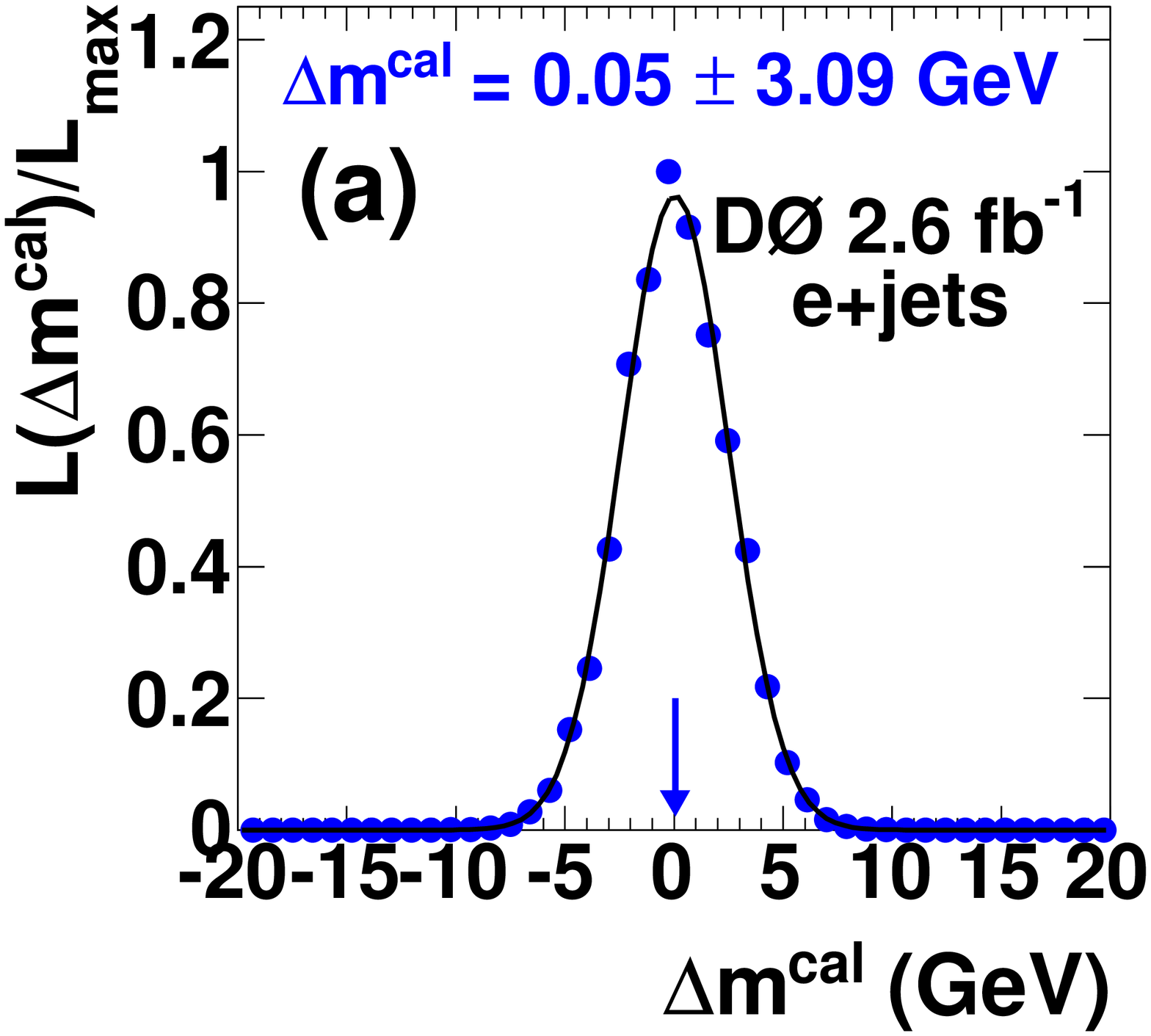}
\hspace{-2mm}
\includegraphics[width=0.24\textwidth]{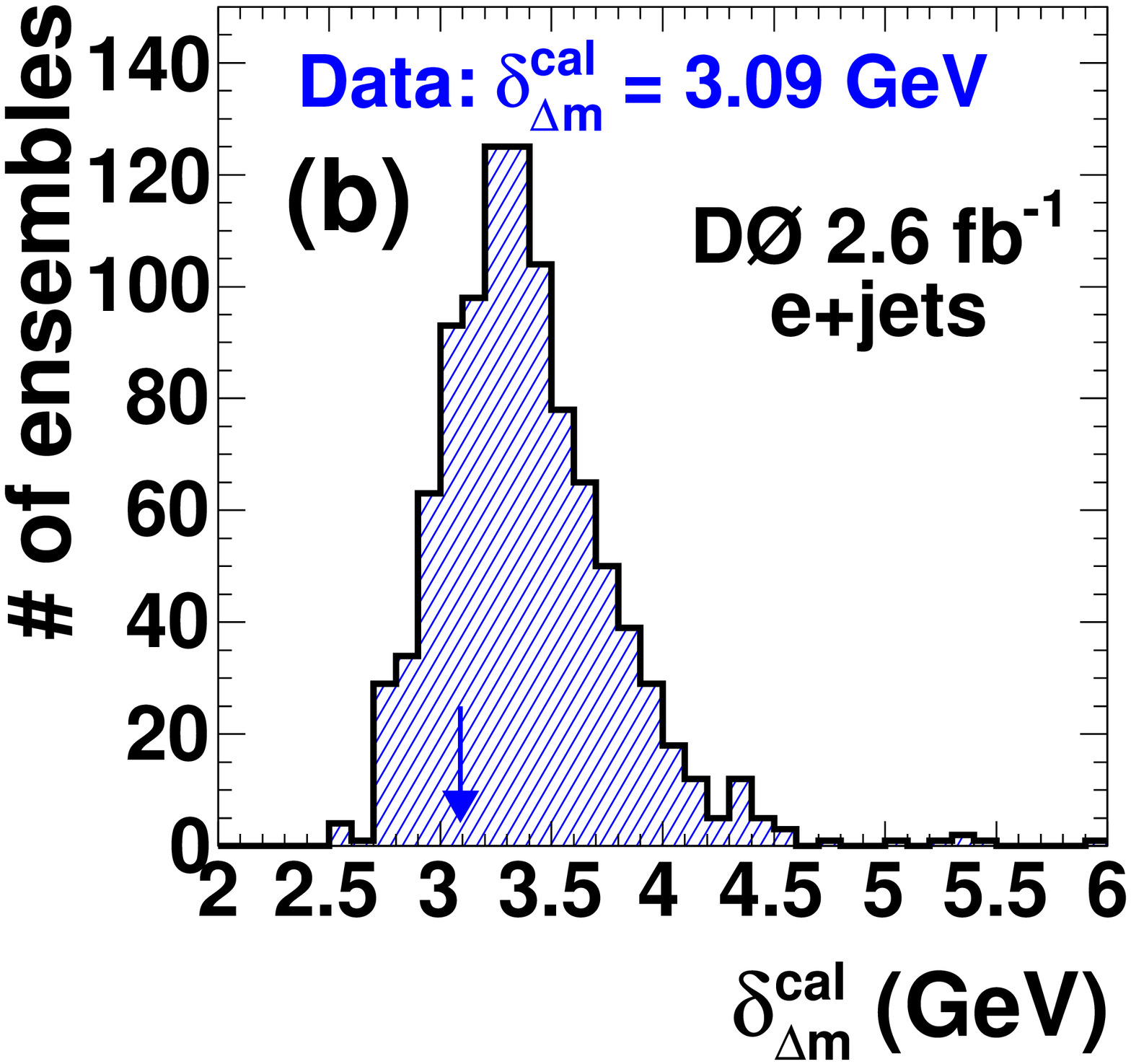}\\
\includegraphics[width=0.24\textwidth]{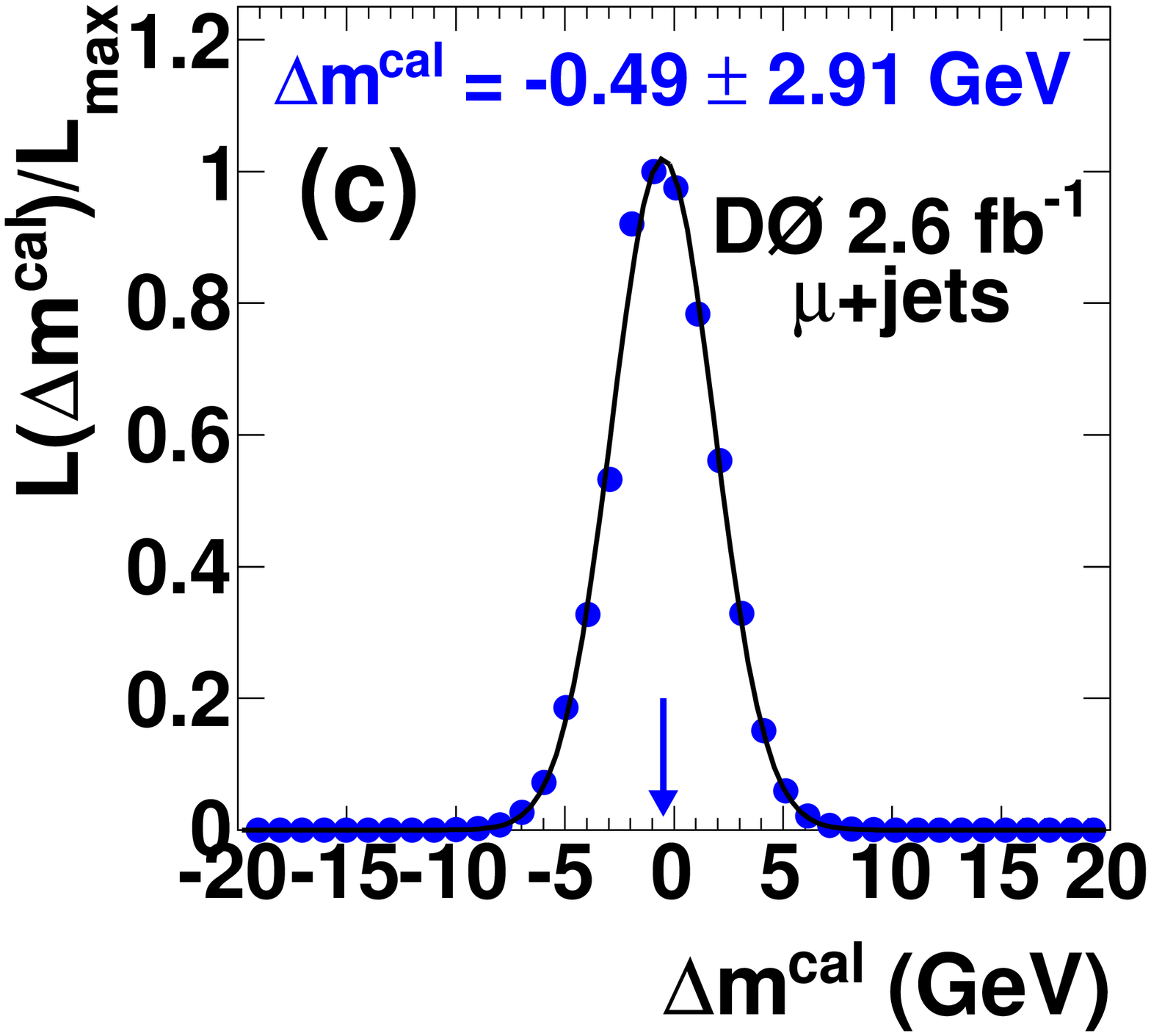}
\hspace{-2mm}
\includegraphics[width=0.24\textwidth]{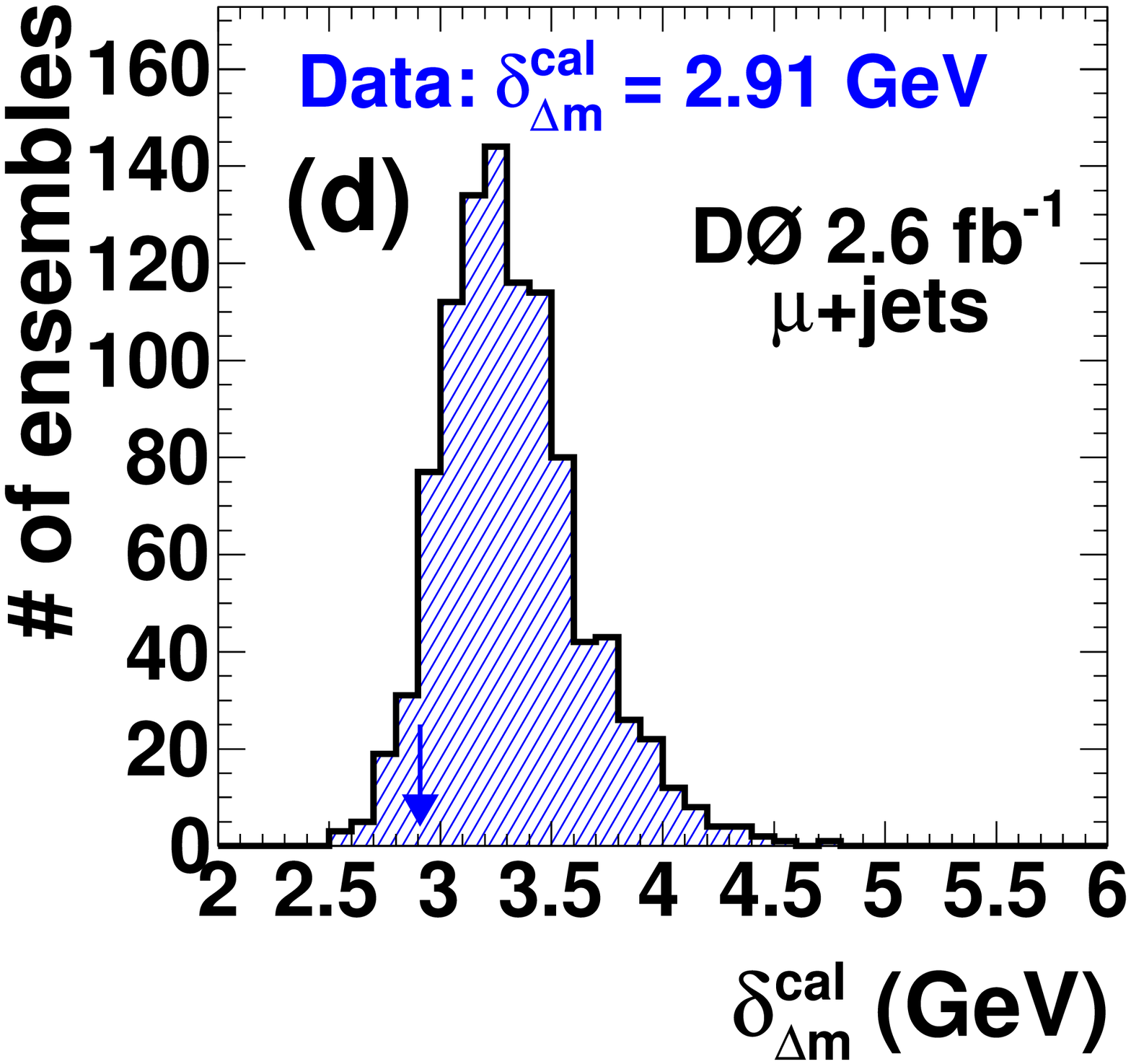}
\caption{
\label{fig:resultsdm}
The normalized likelihood in $\dm^{\rm cal}$ after calibration via Eq.~(\ref{eq:caldm}), together with a Gaussian fit, is shown for the (a)~\ejets and (c)~\mujets events in data. The extracted $\dm^{\rm cal}$ values are indicated by arrows. The distributions in expected uncertainties $\ddm^{\rm cal}$ after calibration via Eq.~(\ref{eq:caldm}) and correction for the pull width, obtained from ensemble studies using simulated MC events, is displayed for the (b)~\ejets and (d)~\mujets channel. The observed $\ddm^{\rm cal}$ values are indicated by arrows.
}
\end{figure}

\begin{figure}
\begin{centering}
\includegraphics[width=0.24\textwidth]{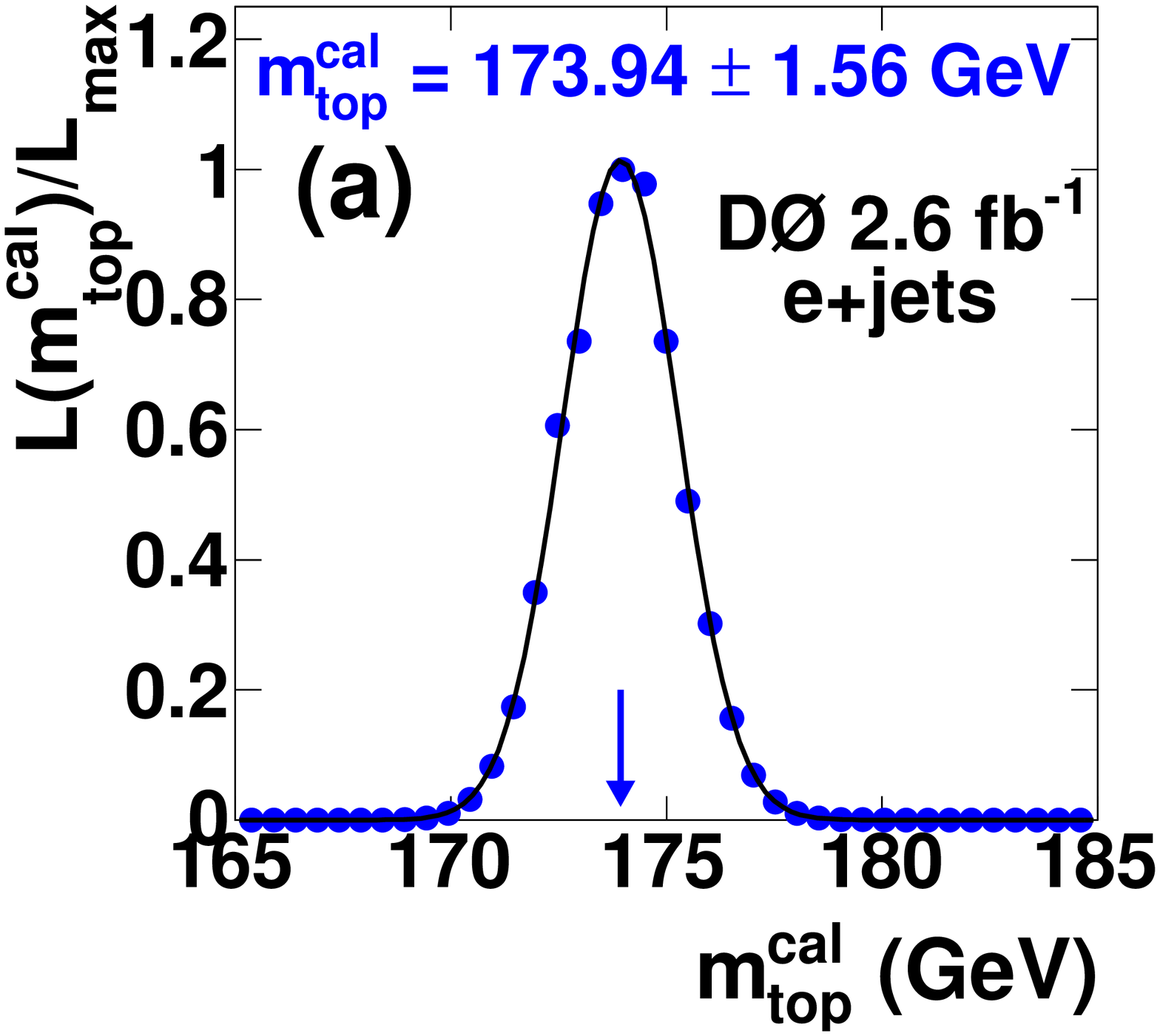}
\hspace{-2mm}
\includegraphics[width=0.24\textwidth]{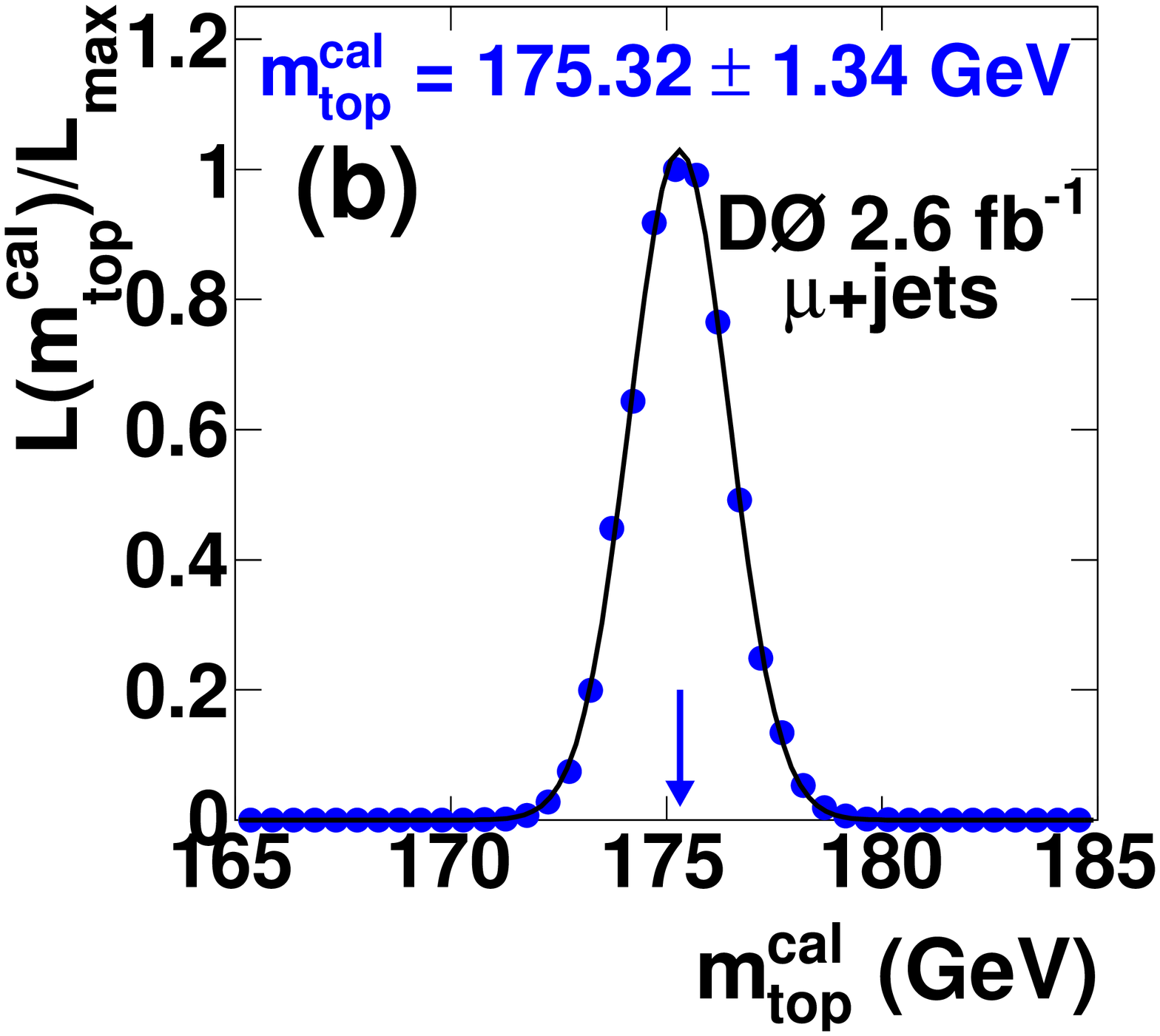}
\end{centering}
\caption{
\label{fig:resultsmsum}
The normalized likelihood in $\msum^{\rm cal}$ after calibration via Eq.~(\ref{eq:calmsum}) together with a Gaussian fit for the (a)~\ejets and (b)~\mujets channel. Arrows indicate the extracted $\msum^{\rm cal}$ values.
}
\vspace{-0.2cm}
\end{figure}

The calibration of data involves a linear transformation of the uncalibrated axes of the likelihoods in $\dm$ and $\msum$ to their corrected values, which we denote as $\dm^{\rm cal}$ and $\msum^{\rm cal}$, according to: 
\begin{eqnarray}
\dm^{\rm cal} &=& \frac{\dm-\xi_{0}^{\dm}}{\xi_{1}^{\dm}},                                 \label{eq:caldm}\\
\msum^{\rm cal} &=& \frac{\msum-172.5~\GeV-\xi_{0}^{\msum}}{\xi_{1}^{\msum}}+172.5~\GeV, \label{eq:calmsum}
\end{eqnarray}
where the $\xi_i$ are summarized in Table~\ref{tab:calib}. The resulting likelihoods for data, as a function of \dm and \msum are shown in Figs.~\ref{fig:resultsdm} and~\ref{fig:resultsmsum}, respectively.

After calibration, $\langle\dm\rangle$ and $\langle\msum\rangle$ with their respective uncertainties $\ddm$ and $\delta_{\msum}$, are extracted from the likelihoods as described in Sec.~\ref{ssec:massfit}. 
The uncertainties are scaled up by the average pull widths given in Table~\ref{tab:calib}. 
The resulting distributions in expected uncertainties $\ddm^{\rm cal}$ are also shown in Fig.~\ref{fig:resultsdm}.
%

The final measured results for \dm and $\msum$ are summarized below according to channel, as well as combined:
\begin{equation}
\begin{array}{rrcrcrl}
\multirow{2}{*}{\ejets,~2.6\,\fb:~~~}                & \dm   & = & 0.1    & \pm & 3.1 & \GeV\,\\
                                                             & \msum & = & 173.9  & \pm & 1.6 & \GeV\,\\
\multirow{2}{*}{\mujets,~2.6\,\fb:~~~} & \dm   & = & -0.5   & \pm & 2.9 & \GeV\,\\
                                                             & \msum & = & 175.3  & \pm & 1.3 & \GeV\,\\
\multirow{2}{*}{\ljets,~2.6\,\fb:~~~}& \dm   & = & -0.2   & \pm & 2.1 & \GeV\,\\
                                                             & \msum & = & 174.7  & \pm & 1.0 & \GeV\,.\\
\end{array}
\label{eq:results}
\end{equation}
The uncertainties given thus far are purely statistical. The combined \ljets results are obtained by using the canonical weighted average formulae assuming Gaussian uncertainties. We cross check the above values for $\msum$ with those obtained from the absolute top quark mass analysis~\cite{bib:me26fb,bib:sample_dep_corr} and find them to be consistent.

%
\begin{figure}
\begin{centering}
\begin{centering}
\includegraphics[width=0.24\textwidth]{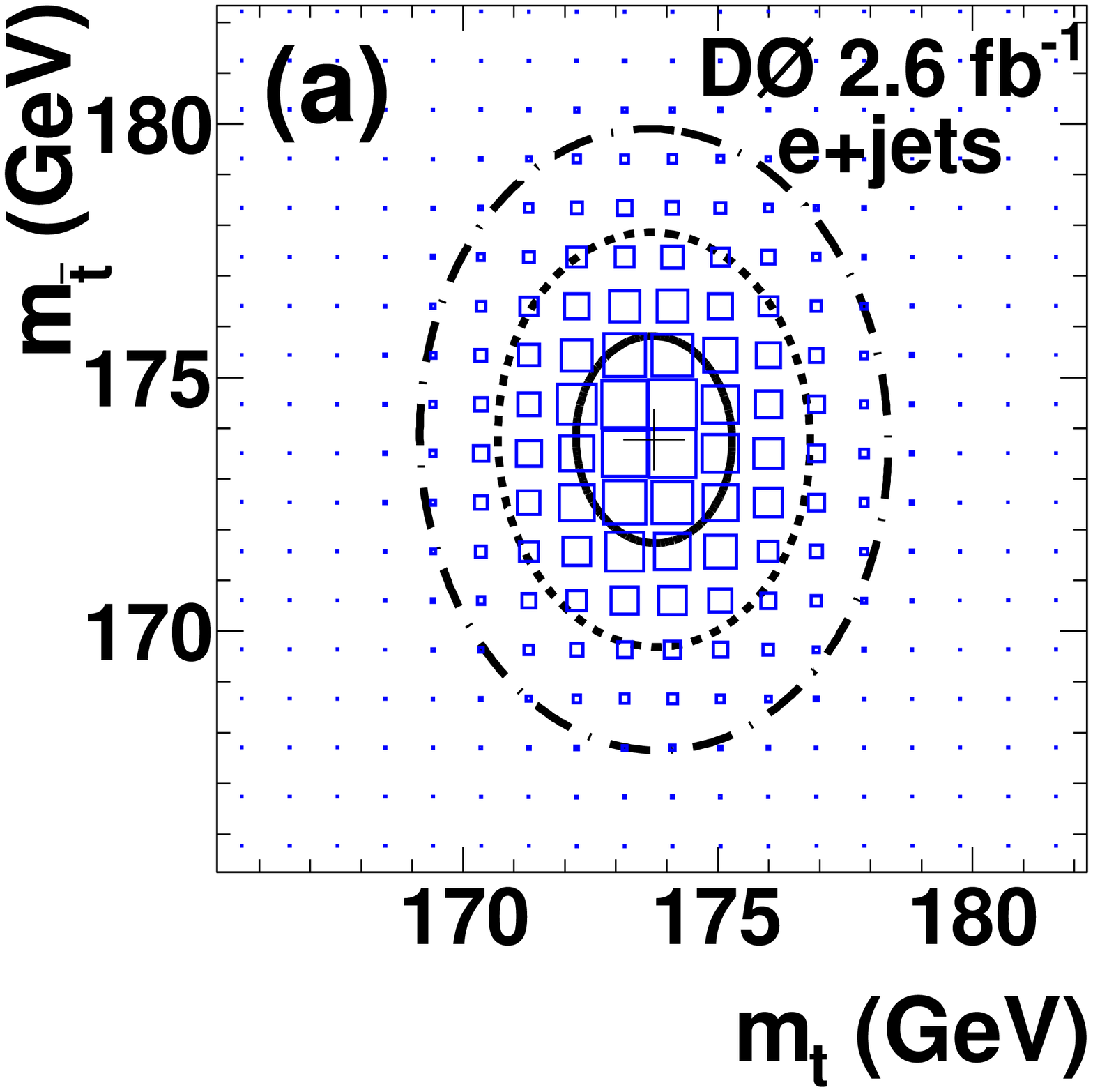}
\hspace{-3mm}
\includegraphics[width=0.24\textwidth]{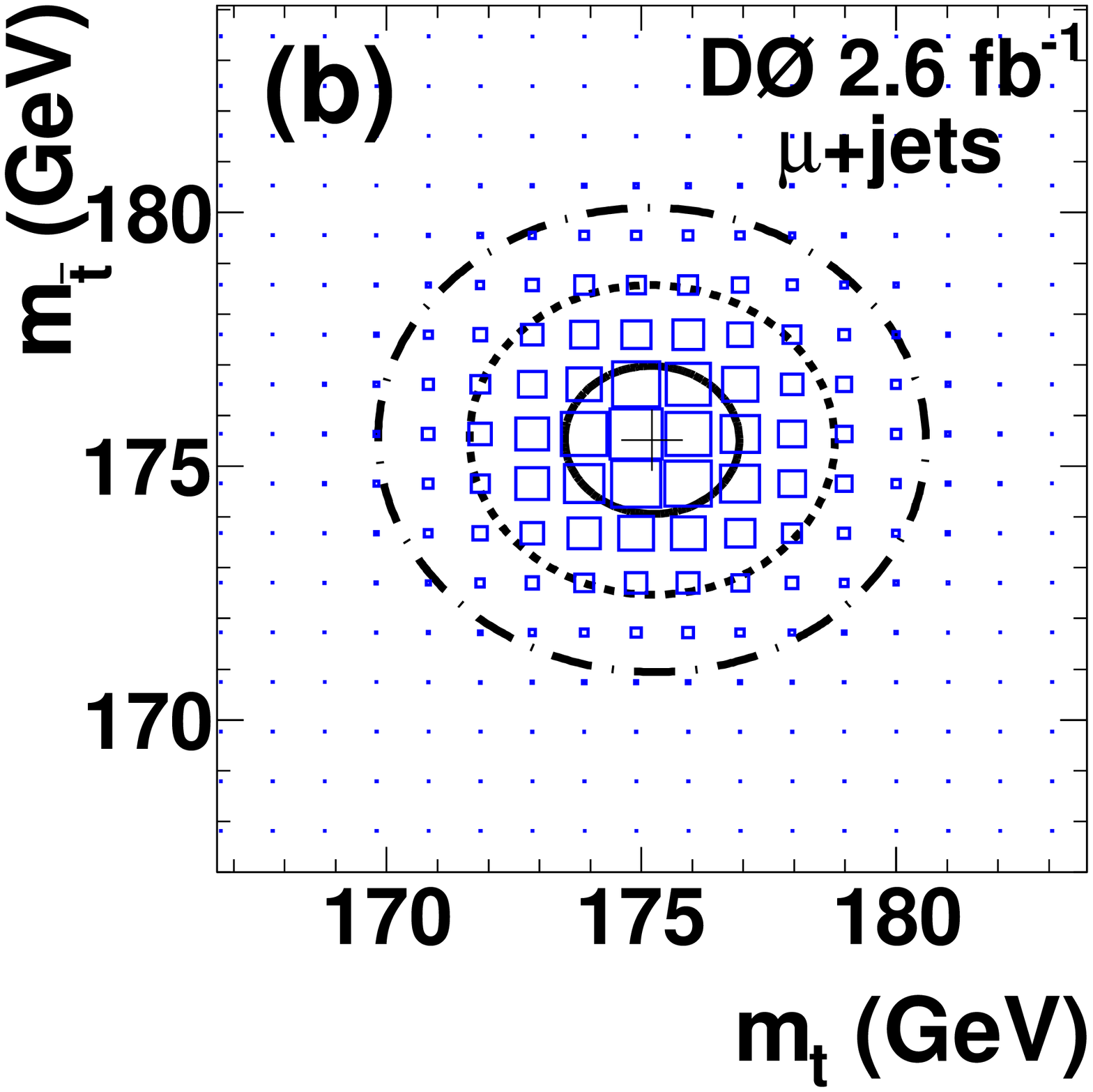}
\end{centering}
\par\end{centering}
\caption{\label{fig:fit2d-emu}
Two-dimensional likelihood densities in $\mt$ and $\mtb$ for the (a)~\ejets and (b)~\mujets channels. The bin contents are proportional to the area of the boxes. The solid, dashed, and dash-dotted lines represent the 1, 2, and 3 SD contours of two-dimensional Gaussian fits (corresponding to approximately 40\%, 90\% and 99\% confidence level, respectively) to the distributions  defined in Eq.~(\ref{eq:gaus}), respectively.
}
\end{figure}
As an additional cross check, we independently extract the masses of the top and antitop quarks from the same data sample. The two-dimensional likelihood densities, as functions of $\mt$ and $\mtb$, are displayed in Fig.~\ref{fig:fit2d-emu}. Also shown are contours of equal probability for two-dimensional Gaussian fits to the likelihood densities, where the Gaussian functions are of the form
\begin{eqnarray}
P(x,y) &=&        \frac{A}{2\pi\sigma_{x}\sigma_{y}}\frac{1}{\sqrt{1-\rho^{2}}} \nonumber\\
        && \times \exp\Big{\{} -\frac{1}{2}\frac{1}{1-\rho^{2}}\Big{[} \frac{(x-\bar{x})^{2}}{\sigma_{x}^{2}} + \frac{(y-\overline{y})^{2}}{\sigma_{y}^{2}} \nonumber\\
        && +      \frac{2\rho(x-\overline{x})(y-\overline{y})}{\sigma_{x}\sigma_{y}}\Big{]}\Big{\}},\label{eq:gaus}
\end{eqnarray}
with $x\equiv \mt$ and $y\equiv \mtb$. The fits to data yield
\begin{equation}
\begin{array}{rrcrcrl}
\multirow{3}{*}{\ejets,~2.6\,\fb:~~~}  & \mt   & = & 173.8  & \pm & 1.5 & \GeV\\
                                       & \mtb  & = & 173.8  & \pm & 2.0 & \GeV\\
                                       & \rho  & = & -0.02\\
\multirow{3}{*}{\mujets,~2.6\,\fb:~~~} & \mt   & = & 175.2  & \pm & 1.8 & \GeV\\
                                       & \mtb  & = & 175.5  & \pm & 1.5 & \GeV\\
                                       & \rho  & = & -0.01.\\
\end{array}
\label{eq:results_2d}
\end{equation}
The above uncertainties are again purely statistical; however, in contrast to Eq.~(\ref{eq:results}), they are not corrected for pull widths in $\mt$ and $\mtb$. The correlation coefficients $\rho$ are consistent with the absence of correlations.

In Sec.~\ref{sec:combi}, we will combine the results for \dm summarized in Eq.~(\ref{eq:results}) with the previous measurement using 1\,\fb of integrated luminosity~\cite{bib:p17dm}.

%

\section{Systematic uncertainties\label{sec:syst}}
For the measurement of \mtop we typically consider
three main types of sources of systematic uncertainties~\cite{bib:me26fb}: 
($i$)~modeling of \ttbar~production and background processes, ($ii$)~modeling of detector response, and ($iii$)~limitations inherent in the measurement method. 
However, in the context of a \dm measurement, many systematic uncertainties are reduced because of correlations between the measured properties of top and antitop quarks, such as, the uncertainty from the absolute JES calibration. Given the small value of the upper limit of $O(5\%)$ already observed for $|\dm|/{\mtop}$, several other  sources of systematic uncertainties relevant in the measurement of \mtop, such as modeling of hadronization, are not expected to contribute to \dm  because they would affect $t$ and $\bar t$ in a similar manner. Following~\cite{bib:barlow}, we check for any effects on \dm that might arise from sources in the latter category in Sec.~\ref{sec:cross}, and find them consistent with having no significant impact. We therefore do not consider them further in the context of this measurement. On the other hand, we estimate systematic uncertainties from additional sources which are not considered in the \mtop measurement, for example from the asymmetry in calorimeter response to $b$ and $\bar b$ quark jets.

Typically, to propagate a systematic uncertainty on some parameter to the final result, that parameter is changed in the simulation used to calibrate the ME method, 
and the \dm result is re-derived. If the change in a parameter can be taken into account through a reweighting of events, a new calibration is determined using those weights and applied directly to data. When this procedure is not possible, a re-evaluation of event probabilities is performed for one sample of \ttbar MC events corresponding to a particular choice of \mt and \mtb closest to the most likely value according to our measurement, i.e. $\mt=\mtb=175~\GeV$, or, when no such sample of MC events with a changed parameter is available, $\mt=\mtb=172.5~\GeV$.
Consequently, the results of ensemble studies are compared to those found for the default sample for the same values of \mt and \mtb.

The systematic uncertainties are described below and summarized in Table~\ref{tab:syst}. 
The total systematic uncertainty is obtained by adding all contributions in quadrature.

\begin{table}
\caption{\label{tab:syst}
Summary of systematic uncertainties on~\dm.}
\begin{centering}
\begin{tabular}{lc}
\hline
\hline
\multirow{2}{*}{Source} & {Uncertainty} \\
 & {on \dm (\GeV)}\\
\hline
Modeling of detector: & \\
\quad Jet energy scale & $0.15$\\
\quad Remaining jet energy scale & $0.05$\\
\quad Response to $b$ and light quarks & $0.09$\\
\quad Response to $b$ and $\bar b$ quarks & $0.23$\\
\quad Response to $c$ and $\bar c$ quarks & $0.11$\\
\quad Jet identification efficiency & $0.03$\\
\quad Jet energy resolution & $0.30$\\
\quad Determination of lepton charge & $0.01$\\
\hline
ME method: & \\
\quad Signal fraction & $0.04$\\
\quad Background from multijet events & $0.04$\\
\quad Calibration of the ME method & $0.18$\\
\hline 
Total & 0.47\\
\hline
\hline
\end{tabular}
\end{centering}
\end{table}

\subsection{Modeling of detector\label{ssec:syst_detector}}
\begin{enumerate}
\item
{\em Jet energy scale:~}
As indicated in Sec.~\ref{ssec:results}, we use the absolute JES calibration of $\jes=1.018\pm0.008$ determined from data. To propagate this uncertainty to \dm, we scale the jet energies in the selected data sample by $\jes\pm1$SD.
\item
{\em Remaining jet energy scale:~}
The systematic uncertainty on the absolute JES discussed above does not account for possible effects from uncertainties on jet energy corrections that depend on $E_{\rm jet}$ and $\eta_{\rm jet}$. To estimate this effect on \dm, we rescale the energies of jets in the default \ttbar MC sample by a differential scale factor $S(E_{\rm jet},\eta_{\rm jet})$ that is a function of the JES uncertainties, but conserves the magnitude of the absolute JES correction.
%
%
\item
{\em Response to $b$ and light quarks:~}
The difference in the hadronic/electromagnetic response of the calorimeter leads to differences in the response to $b$ and light quarks between data and simulation. 
This uncertainty is evaluated by re-scaling the energies of jets matched to $b$ quarks in the default \ttbar MC sample. 
%
%
\item
{\em Response to $b$ and $\bar b$ quarks:~}
The measurement of \dm can be affected by differences in the reconstruction of the transverse momenta of particles and antiparticles. A~difference could in principle be caused by different \pt scales for $\mu^+$ and $\mu^-$. However, the data consist of an almost equal mix of events with opposite magnet polarities, thereby minimizing such biases. We do not observe any difference in calorimeter response to $e^+$ and $e^-$.
\\
{\phantom{123}}A systematic bias to $\dm$ can also be caused by differences in calorimeter response to quarks and antiquarks. In the case of \ttbar events, this bias could arise especially from a different response to $b$ and $\bar b$-quarks. 
Several mechanisms could contribute to this, most notably a different content of $K^+/K^-$ mesons, which have different interaction cross sections. 
In our evaluation of this systematic uncertainty, we assume that, although differences in response to $b/\bar b$ quarks are present in data, they are not modeled in MC events.
We measure the difference of the calorimeter response to $b$ quarks to that of $\bar b$ quarks, $\rbb\equiv\mathcal{R}_{b}-\mathcal{R}_{\bar b}$, using a ``tag-and-probe'' method in data. Namely, we select back-to-back dijet events, and enhance the $b\bar b$ content by requiring $b$-tags for both jets. The tag jet is defined by the presence of a muon within the jet cone, whose charge  serves as an indication whether the probe jet is more likely to be a $b$ or a $\bar b$-quark jet. By evaluating the $|\vec \pt|$ imbalance between tag and probe jets for positively and negatively charged muon tags, we find an upper bound $|\rbb|<0.0042$. Based on this result, we modify the default \ttbar MC sample by re-scaling the momenta $|\vec p|$ of $b$ ($\bar b$)-quark jets by $1\mp\frac12\cdot\rbb=0.9979$~($1.0021$), and adjusting their 4-vectors accordingly. We repeat the ensemble studies after recalculating the probabilities for the modified sample and quote the difference relative to the default sample as a systematic uncertainty.
%
%
%
\item
{\em Response to $c$ and $\bar c$ quarks:~}
A difference in calorimeter response to $c$ and $\bar c$ quarks can potentially bias \dm, since~$c$ quarks appear in decays of $W^+$ bosons from $t$ quark decays, and vice versa for $\bar c$ and $\bar t$. It is expermentally difficult to isolate a sufficiently clean sample of $c\bar c$ dijet events, since it will suffer from considerable contributions from $b\bar b$ dijet events. However, the major underlying mechanisms that could cause a response assymetry, like, e.g., the different content of $K^+/K^-$ mesons, are the same, but of roughly opposite magnitude between $c$ and $b$ quark jets, which would result in an anticorrelation. Based on the above, we assume the same upper bound $|\mathcal{R}_{c,\bar c}|\leq\mathcal{R}_{b,\bar b}<0.0042$, and treat $\mathcal{R}_{c,\bar c}$ and $\mathcal{R}_{b,\bar b}$ as uncorrelated. To propagate the systematic uncertainty from $\mathcal{R}_{c,\bar c}$ to \dm, we apply a similar technique to that for the estimation of the systematic uncertainty due to different response to $b$ and $\bar b$ quarks.
%
%
%
%
%
%
\item
{\em Jet identification efficiency:~}
\dzero uses scale factors to achieve data/MC agreement in jet identification efficiencies. To propagate to the \dm measurement the effect of uncertainties on these scale factors, we decrease the jet identification efficiencies in the default \ttbar sample according to their uncertainties.
%
%
\item
{\em Jet energy resolution:~}
An additional smearing of jet energies derived by comparison of the $\pt$ balance in $(Z\rightarrow ee)+1\,{\rm jet}$ events~\cite{bib:jer} is applied to all MC samples in this analysis in order to achieve better data/MC agreement. To evaluate any effect from data/MC disagreement in jet energy resolutions on~$\dm$, we modify the default \ttbar MC sample by varying the jet energy resolution within its uncertainty. 
%
%
\item
{\em Determination of lepton charge:~}
This analysis uses the charge of the lepton in \ttbar candidate events to distinguish the top quark from the antitop quark. Incorrectly reconstructed lepton charges can result in a systematic shift in the measurement. The charge misidentification rate is found to be less than 1\% in studies of $Z\to ee$ data events. To estimate the contribution of this uncertainty, we assume a charge misidentification rate of 1\% for both \ejets and \mujets final states and evaluate the effects on \dm resulting from a change in the mean values of the extracted $\mt^{\rm cal}$ and $\mtb^{\rm cal}$.
\end{enumerate}

\subsection{ME method}
\begin{enumerate}
\item
{\em Signal fraction:~}
The signal fractions \fsig presented in Table \ref{tab:sfrac} are changed by their respective uncertainties for each decay channel, and ensemble studies are repeated for all MC samples to re-derive the calibration for $\dm$. The new calibrations are applied to data and the results compared with those obtained using the default calibration.
\item
{\em Background from multijet events\label{sub:sysqcd}:~}
In the calibration of this analysis, the background contribution to pseudo-experiments is formed using only $W$+jets events, as they are also assumed to model the small MJ background from QCD processes and smaller contributions from other background processes present in the data. To estimate the systematic uncertainty from this assumption, we define a dedicated MJ-enriched sample of events from data.
The calibration is re-derived with this background sample included in forming pseudo-experiments.
%
%
\item
{\em Calibration of the ME method:~}
The statistical uncertainties associated with the offset~($\xi_0$) and slope~($\xi_1$) parameters that define the mass calibration in Sec.~\ref{ssec:calib} contribute to the uncertainty on \dm. To quantify this, we calculate the uncertainty \ddm due to $\delta_{\xi_0}$ and $\delta_{\xi_1}$  for each channel according to the error propagation formula
\[
~~~~~~~\ddm = \left\{ 
                          \left(\frac{\dm-\xi_0}{\xi_1^2}\cdot\delta_{\xi_1}\right)^2
                        + \left(\frac{\delta_{\xi_0}}{\xi_1}\right)^2
          \right\}^{-\frac12}
\]
and then combine the resulting uncertainties for the \mbox{\ejets} and \mujets channels in quadrature.
\end{enumerate}

\subsection{Additional checks\label{sec:cross}}
We check for effects on \dm from sources of systematic uncertainties considered in the \msum measurement~\cite{bib:me26fb} which are not expected to contribute any bias in the context of the measurement of \dm. For this, we follow the same approach as outlined at the beginning of this Section.
We find the results of our checks to be indeed consistent with no bias on \dm. 

The additional checks are described below and summarized in Table~\ref{tab:cross}. Note that the numbers quoted merely reflect an upper bound on a possible bias, rather than any true effect. 
This limitation is statistical in nature and due to the number of available simulated MC events. Furthermore, if the difference between the central result and the one obtained for a check is smaller than the statistical uncertainty on this difference, we quote the latter.

\begin{table}
\caption{\label{tab:cross}
Summary of additional checks for a possible bias on~\dm. None of those show any significant bias on~\dm. Note that the numbers shown reflect an upper limit on a possible bias. This limitation is of statistical origin and due to the number of available simulated MC events.}
\begin{centering}
\begin{tabular}{lc}
\hline
\hline
\multirow{2}{*}{Source} & {Change in \dm} \\
 & {(\GeV)}\\
\hline
Modeling of physical processes: & \\
\quad Higher-order corrections & $0.26$ \\
\quad ISR/FSR & $0.21$ \\
\quad Hadronization and underlying event & $0.23$ \\
\quad Color reconnection & $0.27$ \\
\quad $b$-fragmentation & $0.03$ \\
\quad PDF uncertainty & $0.10$ \\
\quad Multiple hadron interactions & $0.06$ \\
\quad Modeling of background & $0.07$\\
\quad Heavy-flavor scale factor & $0.02$\\
\hline 
Modeling of detector: & \\
\quad Trigger selection& $0.07$\\
\quad $b$-tagging efficiency & $0.25$\\
\quad Momentum scale for $e$ & $0.05$\\
\quad Momentum scale for $\mu$ & $0.06$\\
\hline 
\hline
\end{tabular}
\end{centering}
\end{table}

\subsubsection{Modeling of physical processes\label{ssec:physics}}
%
\begin{enumerate}
\item
{\em Higher-order corrections:~}
To check the effect of higher-order corrections on~\dm, we perform ensemble studies using \ttbar~events generated with $(i)$~the NLO MC generator {\sc mc@{}nlo}~\cite{bib:mcnlo}, and $(ii)$~the LO MC generator \alpgen, with \herwig~\cite{bib:herwig} for hadronization and shower evolution.
\item
{\em Initial and final-state radiation:~}
The modeling of extra jets from ISR/FSR is checked by comparing \pythia samples with modified input parameters, such as the $\pm1$SD changes, found in a study of Drell-Yan processes~\cite{bib:cdfisr}.
\item
{\em Hadronization and underlying event:~}
To check a possible effect of \dm from the underlying event as well as the hadronization models, we compare samples hadronized using \pythia with those hadronized using \herwig.
\item
{\em Color reconnection:~}
The default \pythia tune used at \dzero (tune {\tt A}), does not include explicit color reconnection. 
For our check, we quantify the difference between \dm values found in ensemble studies for \ttbar MC samples generated using tunes {\tt Apro} and {\tt ACRpro}, where the latter includes an explicit model of color reconnection~\cite{bib:color,bib:acrpro}.
\item
{\em $b$-fragmentation\label{ssub:bfrag}:~}
Uncertainties in the simulation of \mbox{$b$-quark} fragmentation can affect the measurement of $\msum$ in several phases of the analysis, such as in $b$-tagging and in the $b$-quark transfer functions used in the ME calculations. Such effects are studied in the context of \dm by reweighting the simulated $t\bar{t}$ events used in the calibration of the method from the default Bowler scheme~\cite{bowler}, which is tuned to LEP (ALEPH, OPAL, and DELPHI) data, to a tune that accounts for differences between SLD and LEP data~\cite{bfragyvonne}. 
\item
{\em Uncertainty on PDF:~}
The CTEQ6M~\cite{bib:cteq} PDFs provide a set of possible excursions in parameters from their central values. To check the effect on \dm from PDFs, we change the default \ttbar MC sample (generated using CTEQ6L1) by reweighting it to CTEQ6M, repeat the ensemble studies for each of the parameter variations, and evaluate the uncertainty using the prescribed formula~\cite{bib:cteq}:
\[
\qquad\delta_{\dm,\rm PDF}=\frac{1}{2}\bigg\{\sum{}_{i=1}^{20}[\dm(S_{i}^{+})-\dm(S_{i}^{-})]^{2}\bigg\}^\frac12,
\]
where the sum runs over PDF uncertainties for positive ($S_{i}^{+})$ and negative ($S_{i}^{-}$) excursions.
\item
{\em Multiple hadron interactions:~}
When calibrating the ME method, we reweight the luminosity profiles of our MC samples to the instantaneous luminosity profile for that data-taking period. For our check, we re-derive the calibration ignoring luminosity-dependent weights.
\item
{\em Modeling of background:~}
We check the effect of inadequate modeling of background processes on our \dm measurement by identifying distributions in the background-dominated $\ell+3$\,jets events that display only limited agreement between data and predictions from the sum of our signal and background models, as determined through a Kolmogorov-Smirnov test~\cite{bib:kstest}. The calibration of the method is then re-done using \wjets events that are reweighted to bring the identified distributions of predicted signal and background events into better agreement with data. 
%
%
\item
{\em Heavy-flavor scale-factor:~}
As discussed in Sec.~\ref{sec:samples}, a heavy-flavor scale-factor of $1.47\pm0.22$ is applied to the $W\!\!+\!b\bar{b}\!+\!\jets$ and $W\!\!+\!c\bar{c}\!+\!\jets$ production cross sections to increase the heavy-flavor content in the \alpgen $W$+jets MC samples. Moreoever, a scale factor of $1.27\pm0.15$ for the $W\!\!+\!c\!+\!\jets$ production cross section is obtained using {\sc mcfm}.
We re-derive the calibration with the heavy-flavor scale-factor changed by $\pm30\%$ to check the magnitude of the effect on \dm.
\end{enumerate}

\subsubsection{Modeling of detector}
\begin{enumerate}
\item
{\em Trigger selection:~}
To check the magnitude the effect from differential trigger efficiencies on $\dm$, we re-derive a new $\dm$ calibration ignoring the trigger weights.
\item
{\em $b$-tagging efficiency:~}
We check the possibility of a bias in our \dm measurement from discrepancies in the $b$-tagging efficiency between data and MC events by using {\em absolute} uncertainties on the $b$-tagging efficiencies, and account independently for possible discrepancies that are {\em differential} in $\eta$ and \pt of the jet by reweighting the $b$-tagging rate in simulated \ttbar MC events to that observed in data. The total magnitude of a possible effect is determined by combining in quadrature excursions of \dm values obtained with the modified calibrations for both absolute and differential changes.
\item
{\em Momentum scale for electrons:~}
\dzero calibrates the energy of electrons based on studies of the $Z\to ee$ mass for data and MC events.
We rescale the electron energies in the default signal MC sample according to the uncertainties on the electron energy calibration to check the magnitude of the effect in the context of \dm.
%
%
\item
{\em Momentum scale for muons:~}
The absolute momentum scale for muons is obtained from $J/\psi\to\mu\mu$ and $Z\to\mu\mu$ data. However, both linear and quadratic interpolation between these two points can be employed for the calibration.
We check the effect of each extrapolation on \dm by applying the respective corrections to simulated \ttbar MC events in the default sample, and 
find a larger shift in \dm for the linear parametrization.
\end{enumerate}

\section{Combining the 2.6\,${\bf fb}^{\boldsymbol{-1}}$ and 1\,${\bf fb}^{\boldsymbol{-1}}$ analyses\label{sec:combi}}
We use the BLUE method~\cite{bib:blue1,bib:blue2} to combine our new measurement~(Eq.~\ref{eq:results}) with the result of the analysis performed on data corresponding to 1\,\fb~\cite{bib:p17dm}. 
The BLUE method assumes Gaussian uncertainties and accounts for correlations among measurements.\\
For reference, we summarize the results obtained for~1\,\fb:
\[
\begin{array}{rrcrcrl}
\ejets,~1\,\fb:~~~                 & \dm   & = & 0.3   & \pm & 5.0~{\rm(stat)} & \GeV,\\
\mujets,~1\,\fb:~~~  & \dm   & = & 6.7   & \pm & 4.7~{\rm(stat)} & \GeV,\\
\ljets,~1\,\fb:~~~ & \dm   & = & 3.8   & \pm & 3.4~{\rm(stat)} & \GeV.\\
\end{array}
\]


The 1\,\fb analysis used a data-driven method to estimate systematic uncertainties from modeling of signal processes. This method did not distinguish between different sources of systematic uncertainties such as:
$(i)$~higher-order corrections,
$(ii)$~initial and final state radiation,
$(iii)$~hadronization and the underlying event, and 
$(iv)$~color reconnection. 
The above sources are studied in the context of the \msum measurement~\cite{bib:me26fb}, but are not expected to contribute any bias to the measurement of \dm. We cross-check their impact on \dm in Sec.~\ref{sec:cross}, and find them consistent with no bias. Based on our findings, we do not consider any systematic uncertainties from modeling of signal and background processes.

Two sources of systematic uncertainties from modeling of detector peformance (Table~\ref{tab:syst}) are taken to be uncorrelated between the two measurements: JES and remaining JES. The rest are taken to be fully correlated.

In the 1\,\fb analysis, a systematic uncertainty of 0.4~GeV from the difference in calorimeter response to $b$ and $\bar b$ quarks was estimated using MC studies and checks in data. This systematic uncertainty has been re-evaluated using an entirely data-driven approach (item~(iv) in Sec.~\ref{ssec:syst_detector}), and we therefore use this new result for the analysis of the 1\,\fb data. Furthermore, we now evaluated a systematic uncertainty from the difference in calorimeter response to $c$ and $\bar c$ quarks, and propagate our findings to the 1\,\fb analysis.

All other systematic uncertainties not explicitly mentioned above are taken as uncorrelated.

\begin{figure}
\begin{centering}
\includegraphics[width=0.24\textwidth]{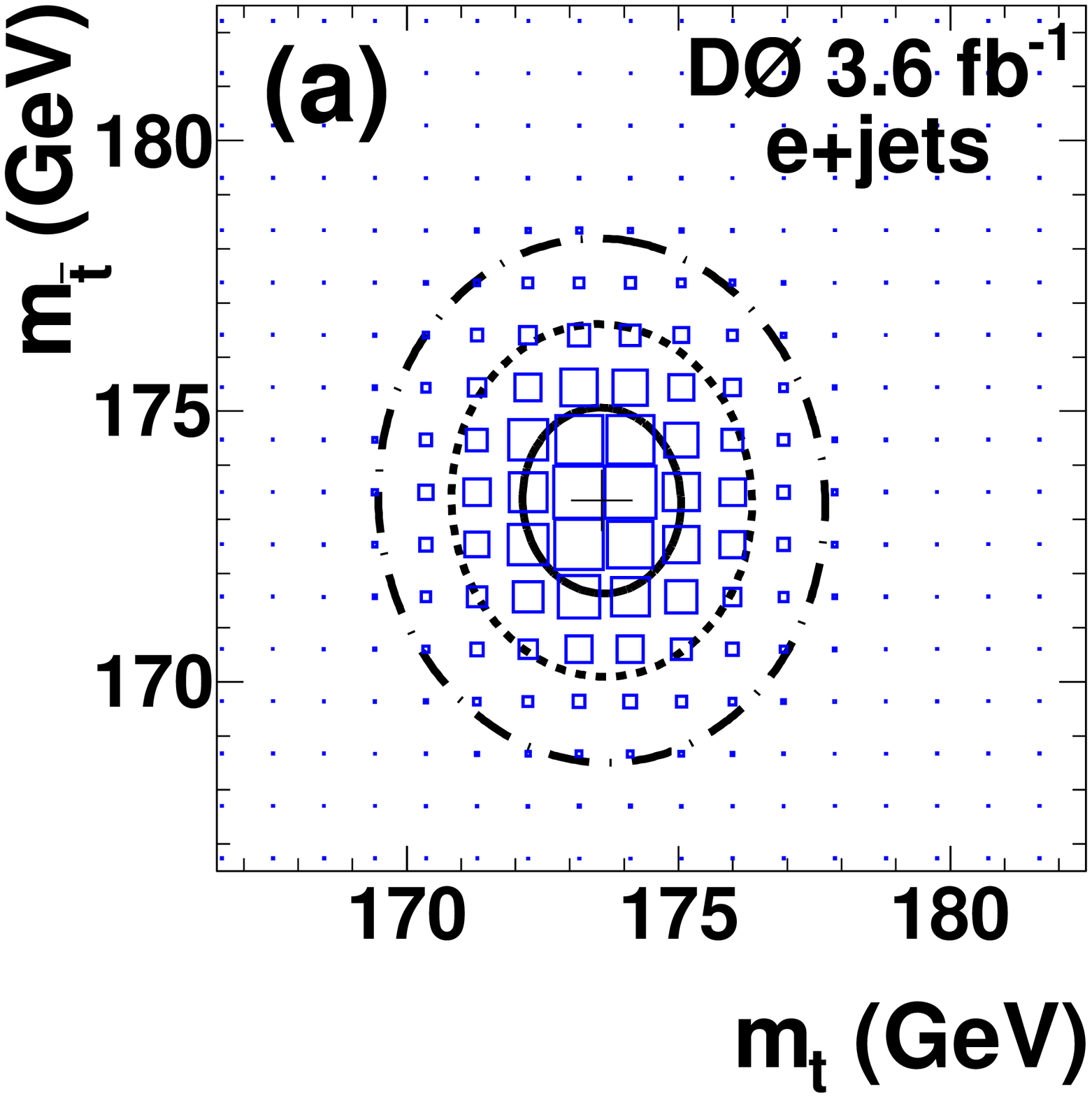}
\hspace{-3mm}
\includegraphics[width=0.24\textwidth]{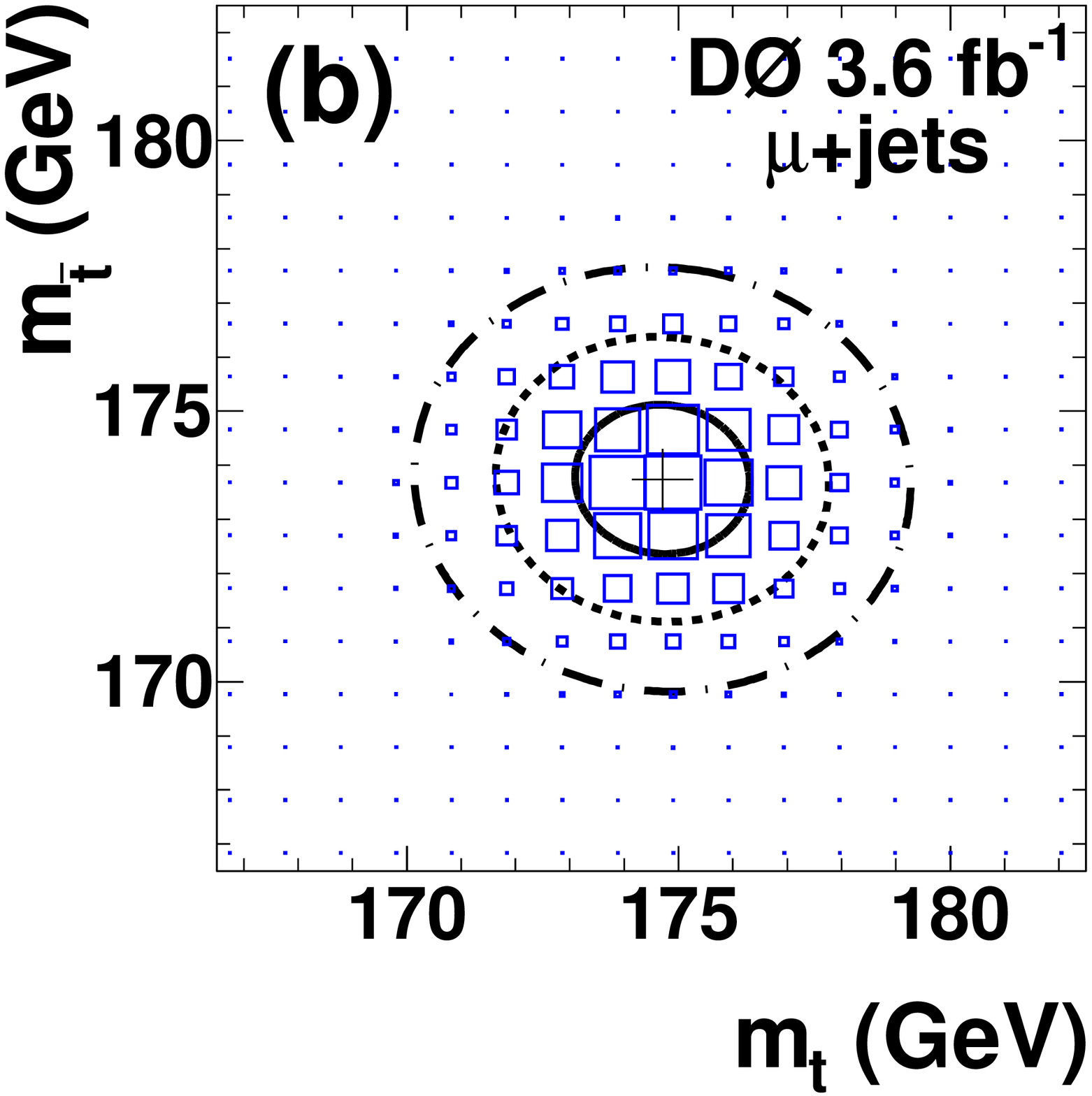}
\end{centering}
\caption{
\label{fig:combiLH2d}
Combined likelihoods of the 2.6\,\fb and 1\,\fb measurements as functions of \mt and \mtb in data for the (a)~\ejets and (b)~\mujets channel. The bin contents are proportional to the area of the boxes. The solid, dashed, and dash-dotted lines represent the 1, 2, and 3 SD contours of two-dimensional Gaussian fits defined in Eq.~(\ref{eq:gaus}) (corresponding to approximately 40\%, 90\% and 99\% confidence level, respectively) to the distributions, respectively. No pull corrections have been applied, and therefore the figures are for illustrative purposes only.
}
\end{figure}

The combined result for \dm corresponding to 3.6\,\fb of data is
\begin{equation}
\dm=0.84\pm1.81~(\rm stat.)\pm0.48~(\rm syst.)~\GeV\,.
\end{equation}
In this combination, BLUE determines a relative weight of 72.8\%~(27.2\%) for the 2.6\,\fb~(1\,\fb) measurement.
The $\chi^2/N_{\rm DOF}$ of the combination is 0.96. The combined likelihood densities for the two analyses are presented in Fig.~\ref{fig:combiLH2d} as functions of \mt and \mtb, separately for the \ejets and \mujets channels.

\section{Conclusion\label{sec:conclusion}}
We have applied the matrix element method to the measurement of the mass difference \dm between top and antitop quarks using $t\bar{t}$ candidate events in the lepton$+$jets channel in data corresponding to an integrated luminosity of about~3.6\,\fb. We find
\[
\dm=0.8\pm1.8~(\rm stat.)\pm0.5~(\rm syst.)~\GeV\,,
\]
which is compatible with no mass difference at the level of $\approx$1\% of the mass of the top quark. 


\setcounter{section}{0}
\section{Appendix: generation of $\boldsymbol{t\bar t}$ events with $\boldsymbol{M_t\neq M_{\boldsymbol{\bar t}}}$\label{sec:app}}
We briefly describe below the modifications to the standard \pythia~\cite{bib:pythia} code which were necessary to generate \ttbar events with $\mt\neq\mtb$.
A new entry in the {\sc kf} particle table is created for the $\bar t$~quark. The {\sc pyinpr} subroutine is modified for use cases in which one of the $\ttbar$ production subprocesses ({\sc isub}${}=81,82,84,85$) is called. The $\bar t$ quark is assigned as the second final-state particle whenever a $t$ quark is selected as the first final-state particle. Furthermore, the ordering of the first and second final-state particles are swapped, as needed, in the subroutine {\sc pyscat}. Additional changes are made in the subroutines {\sc pymaxi}, {\sc pyrand}, and {\sc pyresd} to set the lower limit on the combined masses of the $W^+$ ($W^-$) boson and $b$ ($\bar b$) quark to the $t$ ($\bar t$) quark mass. Finally, the subroutine {\sc pywidt} is modified to adjust the resonance widths $\Gamma_t$ and $\Gamma_{\bar t}$ as functions of \mt and \mtb.

\section*{Acknowledments}
%
We thank the staffs at Fermilab and collaborating institutions,
and acknowledge support from the
DOE and NSF (USA);
CEA and CNRS/IN2P3 (France);
FASI, Rosatom and RFBR (Russia);
CNPq, FAPERJ, FAPESP and FUNDUNESP (Brazil);
DAE and DST (India);
Colciencias (Colombia);
CONACyT (Mexico);
KRF and KOSEF (Korea);
CONICET and UBACyT (Argentina);
FOM (The Netherlands);
STFC and the Royal Society (United Kingdom);
MSMT and GACR (Czech Republic);
CRC Program and NSERC (Canada);
BMBF and DFG (Germany);
SFI (Ireland);
The Swedish Research Council (Sweden);
and
CAS and CNSF (China).
%


\begin{thebibliography}{99}
\bibitem{bib:layer0disclaimer}
The previous measurement in Ref.~\cite{bib:p17dm} is based on data corresponding to 1~\fb of integrated luminosity recorded before the installation of Layer~0, while the new measurement presented here is based on data corresponding to 2.6~\fb of integrated luminosity recorded after this installation.
\bibitem{eta} D0 uses a cylindrical coordinate system with the $z$ axis running 
    along the proton beam axis. The angles $\theta$ and $\phi$ are the polar and azimuthal
    angles, respectively.
    Rapidity $y$ and pseudorapidity $\eta$ are defined as functions
    of the polar angle $\theta$ and parameter $\beta$ as
    $y(\theta,\beta) \equiv
    {\frac{1}{2}} \ln{[(1+\beta\cos{\theta})/(1-\beta\cos{\theta})]}$ and
    $\eta(\theta) \equiv y(\theta,1)$, where $\beta$ is the ratio of a particle's 
    momentum to its energy. Throughout, we define $\eta$ with respect to the center of the
    detector.
\bibitem{ip} The transverse impact parameter is defined as the distance of closest 
    approach (${\rm dca}$) of the track to the PV in the 
    plane transverse to the beamline. Its significance is defined as
    ${\rm dca}/\delta{\rm dca}$, where $\delta{\rm dca}$ is the uncertainty 
    on ${\rm dca}$.
\bibitem{bib:d0tunea}
D0 Tune A is identical to Tune~A~\cite{bib:tunea}, but uses the CTEQ6L1 PDF set and sets $\Lambda_{\rm QCD}=0.165~\GeV$.
\bibitem{bib:sample_dep_corr}
To check the consistency, our measured value of \mtop is to be compared with that found in Ref.~\cite{bib:me26fb}, but before applying the sample-dependent corrections described in Sec.~VII of Ref.~\cite{bib:me26fb}. These corrections produce shifts of $\Delta\mtop = 1.26~\GeV$ and $\Delta k_{\rm JES}=-0.005$. Hence, the values of \mtop from our analysis serve only as cross-checks, whereas the final result in~Ref.~\cite{bib:me26fb} includes all the relevant corrections.
\end{thebibliography}
\end{document}